\documentclass[twocolumn, twocolappendix, times]{aastex631}

\usepackage{newtxmath}
\usepackage{minipage-marginpar}
\usepackage{mathtools}

\usepackage{csquotes}
\usepackage{physics}
\usepackage{natbib}
\usepackage{layouts}
\newcommand{\bcdot}{\boldsymbol{\cdot}}
\usepackage{xpatch}
\usepackage{orcidlink}
\usepackage{xspace}
\usepackage{ulem}
\usepackage{amsmath}

\hypersetup{
	colorlinks=true,
	breaklinks=true,
	allcolors=blue,
	frenchlinks=true
}

\makeatletter
\xpatchcmd\NAT@citex
{%
	\@citea\NAT@hyper@{%
		\NAT@nmfmt{\NAT@nm}%
		\hyper@natlinkbreak{\NAT@aysep\NAT@spacechar}{\@citeb\@extra@b@citeb}%
		\NAT@date
	}%
}
{%
	\@citea
	\NAT@nmfmt{\NAT@nm}%
	\NAT@aysep\NAT@spacechar
	\NAT@hyper@{\NAT@date}%
}
{}{}
\xpatchcmd\NAT@citex
{%
	\@citea\NAT@hyper@{%
		\NAT@nmfmt{\NAT@nm}%
		\hyper@natlinkbreak{\NAT@spacechar\NAT@@open\if*#1*\else#1\NAT@spacechar\fi}%
		{\@citeb\@extra@b@citeb}%
		\NAT@date
	}%
}
{
	\@citea
	\NAT@nmfmt{\NAT@nm}%
	\NAT@spacechar\NAT@@open\if*#1*\else#1\NAT@spacechar\fi
	\NAT@hyper@{\NAT@date}%
}
{}{}
\makeatother

\graphicspath{{./Figs/}}

\newcommand{\ci}{\mathrm{i}} %
\newcommand{\valf}{\ensuremath{\varv_\mathrm{A}}}
\newcommand{\vwa}{\ensuremath{\varv_\mathrm{wave}}}
\newcommand{\vdr}{\ensuremath{\varv_\mathrm{dr}}}
\newcommand{\vperp}{\ensuremath{\varv_\perp}}
\newcommand{\vparwave}{\ensuremath{\varv_\parallel'}}
\newcommand{\mat}[1]{\ensuremath{\textbf{\textsf{#1}}}}

\renewcommand{\vec}[1]{\boldsymbol{#1} }

\newcommand{\btimes}{\boldsymbol{\times}}
\newcommand{\crrm}{\mathrm{cr}}
\newcommand{\Isim}{{\bfseries\ttfamily{I}}\xspace}
\newcommand{\Gfsim}{{\bfseries\ttfamily{F}}\xspace}
\newcommand{\Gbsim}{{\bfseries\ttfamily{B}}\xspace}
\newcommand{\Isimm}{{\boldsymbol{\mathtt{I}}}\xspace}
\newcommand{\Iwsimm}{${\boldsymbol{\mathtt{I_W}}}$\xspace}
\newcommand{\Iesimm}{${\boldsymbol{\mathtt{I_E}}}$\xspace}

\newcommand{\Gfsimm}{{\boldsymbol{\mathtt{F}}} }

\newcommand{\Gsims}{{{\bfseries\ttfamily{F}}\&{\bfseries\ttfamily{B}}}\xspace}

\renewcommand{\phi}{\GenericError} %

\DeclareOldFontCommand{\rm}{\normalfont\rmfamily}{\mathrm}
\DeclareOldFontCommand{\sf}{\normalfont\sffamily}{\mathsf}
\DeclareOldFontCommand{\tt}{\normalfont\ttfamily}{\mathtt}
\DeclareOldFontCommand{\bf}{\normalfont\bfseries}{\mathbf}
\DeclareOldFontCommand{\it}{\normalfont\itshape}{\mathit}
\DeclareOldFontCommand{\sl}{\normalfont\slshape}{\@nomath\sl}
\DeclareOldFontCommand{\sc}{\normalfont\scshape}{\@nomath\sc}

\begin{document}
\shortauthors{Lemmerz et al.}
\shorttitle{The theory of resonant cosmic ray-driven instabilities}

\title{\Large The theory of resonant cosmic ray-driven instabilities -- Growth and saturation of single modes}
\correspondingauthor{Rouven Lemmerz}
\author[0000-0002-4683-8517]{Rouven Lemmerz}
\email{rlemmerz@aip.de}
\affiliation{Leibniz-Institut f{\"u}r Astrophysik Potsdam (AIP), An der Sternwarte 16, 14482 Potsdam, Germany}
\affiliation{University of Potsdam, Institute of Physics and Astronomy, Karl-Liebknecht-Str. 24-25, 14476 Potsdam, Germany}
\author[0000-0001-9625-5929]{Mohamad Shalaby}
\author[0000-0002-7275-3998]{Christoph Pfrommer}
\author[0000-0002-7443-8377]{Timon Thomas}
\affiliation{Leibniz-Institut f{\"u}r Astrophysik Potsdam (AIP), An der Sternwarte 16, 14482 Potsdam, Germany}

	\date{Accepted XXX. Received YYY; in original form ZZZ}

\begin{abstract}
Cosmic ray (CR) feedback is critical for galaxy formation as CRs drive galactic winds, regularize star formation in galaxies, and escape from active galactic nuclei to heat the cooling cores of galaxy clusters. The feedback strength of CRs depends on their coupling to the background plasma and, as such, on the effective CR transport speed. Traditionally, this has been hypothesized to depend on the balance between wave growth of CR-driven instabilities and their damping. Here, we study the physics of CR-driven instabilities from first principles, starting from a gyrotropic distribution of CR ions that stream along a background magnetic field. We develop a theory of the underlying processes that organize the particles' orbits and in particular their gyrophases, which provides an intuitive physical picture of (i) wave growth as the CR gyrophases start to bunch up lopsidedly towards the local wave magnetic field, (ii) instability saturation as a result of CRs overtaking the wave and damping its amplitude without isotropizing CRs in the wave frame, and (iii) CR back-reaction onto the unstable plasma waves as the CR gyrophases follow a pendulum motion around the wave magnetic field. Using our novel fluid-particle-in-cell code fluid-SHARP, we validate our theory on the evolution and excitation of individual unstable modes, such as forward and backward propagating Alfv\'en and whistler waves. We show that these kinetic simulations support our theoretical considerations, thus potentially foreshadowing a fundamental revision of the theory of CR transport in galaxies and galaxy clusters.
\end{abstract}

\section{Introduction}

\subsection{Astrophysical motivation}
Stellar feedback drives galactic winds, which is crucial for understanding the underpinnings of galaxy formation, most prominently the declining star conversion efficiency of gas from the scale of Milky Way-sized galaxies towards dwarf galaxies \citep{Moster2013}. Several physical processes have been suggested to drive those winds: energy and momentum deposition by exploding supernovae can self-regulate the interstellar medium and drive galactic fountains \citep{Simpson2016,Girichidis2016,Girichidis2018,Kim2018}; however, they are not able to drive multiphase galactic outflows to large heights \citep{thomasWhyAreThermally2024}. Ultraviolet radiation emitted by young stellar populations photoionizes the molecular environment and pushes on the gas via radiation pressure, which opens up channels in the optically thick, gas-enshrouding regions, enabling star formation. Radiation can then escape along those channels without providing much feedback \citep{Rosdahl2015}. 

By contrast, cosmic rays (CRs) have long cooling times and dominate the pressure budget in the nearby interstellar medium \citep{Boulares1990}, making a strong case for efficient feedback \citep[see][for a review]{RuszkowskiPfrommer2023}. CRs stream and diffuse through the galaxy to build up an extended pressure distribution from the disk into the galactic halo. As they are advected by galactic outflows above the disk, CRs gradually deposit momentum and energy via wave-particle interactions far from their generation sites, thereby re-energizing and further accelerating galactic winds that can reach out to the virial radius of galactic halos \citep{Uhlig2012,Booth2013,Salem2014,Pakmor2016,Ruszkowski_CRwinds2017,Thomas2023}. This may even cause CRs to dominate the pressure budget in the inner circumgalactic medium, which facilitates the formation of a colder and smoother thermal plasma and has dramatic consequences for the transport of angular momentum of the accreting gas onto the galactic disks and the spatial extents of stellar disks that form from the gaseous phase \citep{Buck2020,Ji2020}. In the cores of galaxy clusters, CRs escape the lobes of AGN jets, and can heat the surrounding cooling plasma to mitigate the cooling-induced collapse and star formation \citep{Guo2008,Pfrommer2013,Ruszkowski2017,Jacob2017a,Jacob2017b}.

\subsection{CR transport and CR-driven instabilities}
To make progress, it is crucial to better understand the physics of CR transport in galaxies and clusters.
The CR streaming instability plays a critical role in the interaction between CRs and their surrounding medium.
In the classic picture set forth by \citet{kulsrudEffectWaveParticleInteractions1969}, this instability hinges on the interplay between CRs driving resonant waves and scattering off of these self-induced waves.
The framework of quasi-linear theory has been foundational in estimating the CR scattering frequency \citep{jokipiiCosmicRayPropagationCharged1966,wentzelPropagationAnisotropyCosmic1969,skillingCosmicRaysGalaxy1971, schlickeiserQuasiLinearTheory1998}.
Quasi-linear theory of CR transport is a perturbation method concerned with evolving a first-order fluctuation (the perturbed CR distribution) around a prescribed ground state (the gyrotropic CR distribution). 
In order to simplify statistical analysis, the random phase approximation is typically employed, which assumes that the rotational phases of the waves and particles are random and uncorrelated.

The scattering frequency of CRs is closely linked to the intensity of the saturated waves, and thus the saturation mechanism is of particular importance.
The amplitude of the resonant waves can saturate as a result of the competition between wave growth and damping processes.
Viable damping processes are thought to include wave damping through ion-neutral collisions \citep{kulsrudEffectWaveParticleInteractions1969,zweibelConfinementCosmicRays1982,ivlevPenetrationCosmicRays2018}, collisionless nonlinear Landau damping \citep{kulsrudEffectWaveParticleInteractions1969, volkNonlinearLandauDamping1982} and turbulent damping \citep{farmerWaveDampingMagnetohydrodynamic2004,Lazarian2016,Lazarian2022,cerriRevisitingRoleCosmicray2024}. Provided there is sufficient wave energy available, CR scatter frequently and isotropize in the wave frame and thus, stream at the wave velocity.

Alternatively, CRs are thought to align their gyrophases with the self-induced wave \citep{briceExplanationTriggeredVerylowfrequency1963, sudanTheoryTriggeredVLF1971}. In this paper we further explore this mechanism and show, that it is an integral part of explaining the instability growth and saturation of the CR streaming instability for single-mode excitation.

Computational advances in recent years have allowed studying the gyroresonant CR streaming instability using particle-in-cell (PIC) simulations \citep{holcombGrowthSaturationGyroresonant2019, shalabyNewCosmicRaydriven2020}, which follow the orbits of macro particles representing individual particles of a plasma, which are subject to electromagnetic fields that obey Maxwell's equations. Alternatively, this can be done with hybrid-PIC \citep{Weidl2019,haggertyHybridSimulationsResonant2019, schroerCosmicrayGeneratedBubbles2022} methods, in which the electron timescale is integrated out, representing the electron population as an adiabatic fluid while treating the ions as macro particles within the kinetic PIC model.

The large scale separation inherent to the streaming problem has also led to the development of new methods, such as MHD-PIC \citep[][]{baiMagnetohydrodynamicParticleinCellMethodCoupling2015,baiMagnetohydrodynamicParticleincellSimulations2019, lebigaKineticMHDSimulations2018,sunMagnetohydrodynamicparticleincellModuleAthena2023}, which has been used to study ion-neutral damping \citep{plotnikovInfluenceIonNeutralDamping2021,baiFirstprincipleCharacterizationCosmicray2022, bambicMHDPICSimulationsCosmicRay2021}.
MHD-PIC describes the thermal plasma using the magnetohydrodynamic (MHD) approximation while capturing the kinetic physics of the CRs using the PIC method. 
This method has been commonly applied together with a scheme to randomize the CR gyrophases, which enforces the random-phase approximation inherent to the theoretical framework of quasi-linear theory.
More recently, the fluid-PIC method \citep{lemmerzCouplingMultifluidDynamics2024} has been devised, which treats the thermal plasma in the warm plasma approximation and which is a more accurate representation of the plasma in comparison to MHD. 
In addition to accurately representing physics at the ion skin-depth, this method can also correctly capture gyroresonant streaming instabilities at scales smaller than the ion skin-depth and also allows emulating nonlinear Landau damping.

\subsection{Idea to elucidate the physics of CR-driven instabilities}

\begin{figure*}
    \centering
    \plotone{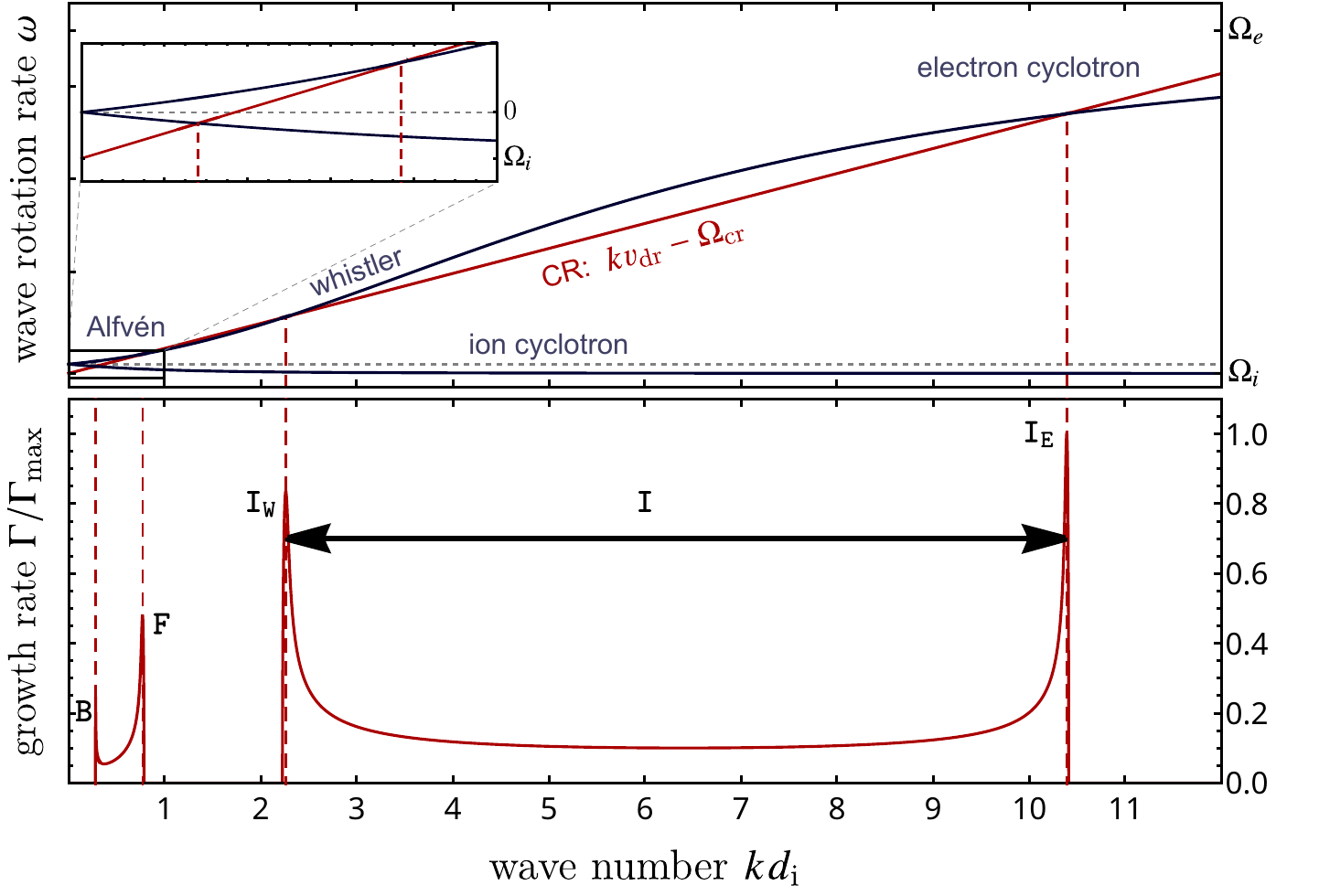}
    \caption{
    CRs drifting and gyrating along the magnetic field can resonantly excite unstable wave modes at different scales. The top panel shows the rotation rate of the CR ion cyclotron mode (red color) as well as the normal modes in the background (black), ranging from forward moving Alfv\'en waves to whistlers to electron cyclotron waves on the upper branch and on the bottom branch, they change character from backward moving Alfv\'en waves to ion cyclotron waves on small scales. The bottom panel shows the instability growth rate of the CR ion cyclotron wave that is maximized at the points of resonances with the background modes. We obtain the solutions by solving the dispersion relation of drifting CRs in a high-density background (equation~\ref{eq:fulldisp-bg}) and evaluate them in the background rest frame. Solely for visualization purposes, we choose parameters that yield comparable grow rates and only a small scale separation: $n_\mathrm{cr}=10^{-6}n_\mathrm{bg}$, $\valf=10^{-4}c$, $m_r=36$, $\varv_\perp=\valf$, and $\vdr=2.7\valf<\valf \sqrt{m_r}/2$. This choice fulfills the condition for exciting the intermediate-scale instability \citep{Shalaby2023}.
    }
    \label{fig:overviewInstability}
\end{figure*}

In this work, we attempt to gain an intuitive picture of the growth and saturation of the resonant CR streaming instability by adopting an innovative and unique approach of isolating the physics of wave growth and wave interactions when studying the saturation behavior. To this end, we run three different simulations, each tailored to excite a single unstable mode at different spatial scales in the background with the goal to understand the essential physics of the instabilities. Specifically, CRs drifting at a mean velocity $\varv_\mathrm{dr}$ along the mean magnetic field of strength $B_0$ can excite forward propagating Alfv\'en waves (denoted by \Gfsim), backward propagating Alfv\'en waves (denoted by \Gbsim), both via the CR streaming instability \citep{Kulsrud1969}, and whistler and electron cyclotron waves via the intermediate-scale instability (denoted by \Iwsimm and \Iesimm, and collectively denoted by \Isim,\ see \citealp{shalabyNewCosmicRaydriven2020,Shalaby2023}). This is visualized in  Figure~\ref{fig:overviewInstability}, which shows the wave rotation rates (top) and growth rates of unstable modes that result from the interaction of CR and background modes (bottom).
CRs transfer energy fastest to the background modes at those wave numbers $k$ at which the rotation rate of the CR ion cyclotron mode, $\omega=k \varv_\mathrm{dr} - \Omega_\mathrm{cr}$ matches the rotation rate of the background  modes, which get modified as a result of the CR-wave interaction (as we will show later in this work).
Here, $\Omega_\mathrm{cr}=qB_0/(\gamma m)$ is the relativistic gyro frequency of a particle of charge $q$, mass $m$, and Lorentz factor $\gamma$. While this explains the possibility for instability growth in the linear regime, here we will specifically address the processes causing linear growth and nonlinear saturation of single wave modes.

In future work, we will study extensions of this picture arising from 1.\ interacting wave modes, 2.\ varying CR-to-background density ratios and Alfv\'en speeds, and 3.\ varying the CR energy and pitch angle distributions, where the pitch angle is measured between an individual CR momentum and the mean magnetic field. We acknowledge that our idealized approach of restricting ourselves to the growth of single wave modes does not necessarily capture the full physics of power-law distributed CRs. However, this enables us to fully understand the underlying physics in this simple setup and to construct an analytic model for the feedback loop, which explains the wave growth, as well as the overall interplay of CRs with waves at the resonance.

The paper is structured as follows. We first introduce our numerical method and setup in Section~\ref{sec:numsetup}, which is followed by theoretical considerations about CR particle orbits and derivation of the pendulum equation for CR-wave interactions in Section~\ref{sec:theory}.
We explain the microphysical mechanism for the linear wave growth in Section~\ref{sec:linear_growth}, while the nonlinear phase of the instabilities and wave saturation is discussed in Section~\ref{sec:nonlinear}.
We conclude our paper in Section~\ref{sec:conclusions}.
We discuss our conventions and compare them to other popular choices in Appendix~\ref{app:convention}. To address the accuracy of our method, we compare a fluid-PIC and a PIC simulation of the intermediate-scale instability in Appendix~\ref{app:PIC} and discuss the solution to the dispersion relation in the background frame in Appendix~\ref{sec:DR}.
Throughout this work, we use the SI system of units. 

\section{Numerical Method and Setup}
\label{sec:numsetup}
In this section, we describe the numerical method and clarify our specific setups and choices for our parameters.

\subsection{Method}

We use the fluid-PIC code fluid-SHARP \citep{shalabySHARPSpatiallyHigherorder2017,shalabyNewCosmicRaydriven2020,lemmerzCouplingMultifluidDynamics2024}, which is an advantageous method for simulating energetic particle transport in a much denser background plasma in comparison to the pure PIC method. The fluid-PIC method combines a hydrodynamic solver, which allows us to treat background particles as a computationally cheap fluid, with a PIC solver that integrates the individual orbits of the energetic particles in the fully kinetic picture.
Both components, the background and energetic particles are coupled via Maxwell's equations.
Here, we give a brief overview of this method.

The CR particles are treated by the SHARP PIC code \citep{shalabySHARPSpatiallyHigherorder2017,shalabyNewCosmicRaydriven2020}, which advances macroparticles that represent CR ions and electrons in one spatial and three velocity dimensions. Moving charges generate currents that induce electromagnetic fields according to Maxwell’s equations. These electromagnetic fluctuations create Lorentz forces that accelerate charged particles, altering the charge distribution and currents. The PIC method evolves this system by numerically iterating this loop on a fraction of the electron plasma timescale, thereby self-consistently taking micro-instabilities driven by these particles into account. 

The more numerous background particles would result in a large computational cost if they were to be treated kinetically, but because they are not driving the instability, they can instead be approximated as a thermal fluid composed of electrons and protons.
This corresponds to the \enquote{warm plasma} model, according to the definition found in many textbooks such as \citet{stixWavesPlasmas1992}, which naturally captures Alfv\'en, whistler, electron cyclotron, ion cyclotron and Langmuir and ion acoustic waves. 
As such, CRs can resonate with the waves carried by the fluids and thus trigger resonant streaming instabilities.
For convenience, we quote the fluid equations solved by the fluid-SHARP code \citep{lemmerzCouplingMultifluidDynamics2024}, which are the fluid continuity, momentum, and energy conservation equations:
\begin{align}
	\frac{\partial n}{\partial t} + \vnabla \bcdot \left(n  \vec{\varw}  \right) &= 0, \label{eq:mass_conservative}\\
	\frac{\partial n\vec{\varw}  }{\partial t} + \vnabla \bcdot \left[ p \mat{1} + n \vec{\varw} \vec{\varw} \right]&= \frac{q}{m} \vec{S}_\varw\left(n, \vec{\varw}, \vec{B}, \vec{E}\right), \label{eq:mom_conservative}\\
	\frac{\partial \epsilon}{\partial t} + \vnabla \bcdot \left[ (p  + \epsilon ) \vec{\varw}  \right] + \frac{1}{\Gamma - 1} \vnabla \bcdot \vec{Q}&= \frac{q}{m} \vec{\varw}\bcdot\vec{S}_\varw\left(n, \vec{\varw}, \vec{B}, \vec{E}\right).
	\label{eq:cons_energy}
\end{align}
The number density is denoted by $n$, the bulk velocity is $\vec\varw$ and the energy and pressure are $\epsilon$ and $p$, respectively. These are evolved for both ion and electron background species separately, which are each characterized by the charge $q$ and particle mass $m$. 
The dyadic product of the two vectors is $\vec{\varw}\vec{\varw}$ and the unit matrix is denoted by $\mat{1}$, indicating an isotropic pressure tensor of the background species.
The energy density and pressure of thermal protons and electrons are separably coupled via the adiabatic index $\Gamma_\mathrm{ad}=5/3$:
\begin{align}
	\epsilon = \frac{p}{\Gamma_\mathrm{ad} - 1} + \frac{1}{2}\,n \,\vec{\varw} \bcdot \vec{\varw}.
\end{align}
Maxwell's equations are used to solve for the electric and magnetic fields $\vec E$ and $\vec{B}$, which exert a force on the fluid that is captured by the source term,
\begin{equation}
	\vec{S}_\varw\left(n, \vec{\varw}, \vec{B}, \vec{E}\right) = n  \left(\vec{E} + \vec{\varw} \btimes \vec{B}\right).
\end{equation}

Even though collisionless physics is not modeled from first principles in this fluid model, it can still be approximated by using appropriate closures.
We use a Landau closure, which models electrostatic Landau damping through a non-local approximation of the heat flux $\vec{Q}$. For further details, we refer the reader to \citet{lemmerzCouplingMultifluidDynamics2024} or the notes by \citet{hunanaIntroductoryGuideFluid2019}.
As we will demonstrate below, in our setups we intentionally only excite individual wave modes, implying that the interference between different modes, and thus the impact of nonlinear Landau damping, is minimized.

\subsection{Setup}
\label{sec:setup}
\begin{figure}
   \resizebox{\hsize}{!} { \includegraphics[width=\textwidth]{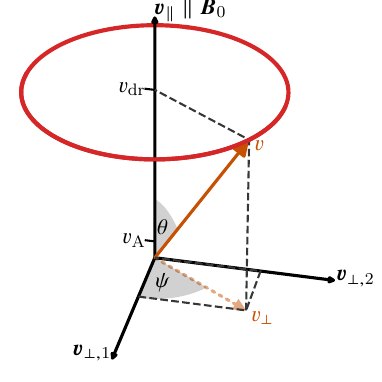} }
	\caption{Visualization of the geometry of our initial CR distribution in velocity space. Two angles are defined, the pitch-angle $\cos(\theta)=\mu=\varv_\parallel/\varv$ and the rotational angle $\tan(\psi)=(\varv_{\perp,2}/\varv_{\perp,1})$.   The CR ion initial conditions are shown as a red circle, with fixed $\theta$ and uniformly distributed $\psi$. A neutralizing CR electron beam at the same $\vdr$ but with $\varv_\perp=0$ is initialized as well.}
	\label{fig:gyro_init}
\end{figure}

CRs are naturally distributed over space, velocity and time, $f(\vec{x}, \vec{\varv}, t)$. In the following, we work in a coordinate system where one of the coordinate axes is aligned with the direction of the static background magnetic field $\vec{B}_0$. Particles that are gyrating because of this magnetic field have rotating velocity components that lie in the plane perpendicular to $\vec{B}_0$. We denote the rotation phase of this gyration by $\psi$. The full particle velocity vector $\vec{\varv}$ further depends on the velocity magnitude $\varv$ and the pitch angle $\theta$. These definitions completely describe our phase space geometry, which we depicted in Figure~\ref{fig:gyro_init}. The presence of the background magnetic field naturally introduces a decomposition of the velocity vector into a parallel component $\varv_\parallel = \mu \varv$ and a perpendicular component $\varv_\perp = \sqrt{1 - \mu^2} \varv$. 

We investigate the interplay of charged particles with transversal waves, which have magnetic field components that also rotate in the plane perpendicular to the background magnetic field $\vec{B}_0$. With no additional information about the distribution of particles in the perpendicular plane, it is customary to assume that all particles are distributed uniformly in rotation angle $\psi$ because in the absence of any transversal magnetic fields there is no distinct direction in the perpendicular plane which could function as a reference direction. We will show that the presence of transversal magnetic fields introduce such a reference direction which ultimately break symmetry and cause anisotropic CR distributions in $\psi$.

Here, we perform three simulations, showcasing instabilities at different scales. 
We study the action of the gyroresonant instability \citep{Kulsrud1969}, which excites Alfv\'en waves at scales larger than the ion skin depth. This instability is further separated into a forward (\Gfsim) and backward (\Gbsim) moving wave, as illustrated in  Figure~\ref{fig:overviewInstability}.
In addition, we will study the intermediate-scale instability \citep[\Isim,][]{shalabyNewCosmicRaydriven2020,Shalaby2023}, which excites whistler waves (\Iwsimm) and electron cyclotron waves (\Iesimm) below the scale of the ion skin depth (see  Figure~\ref{fig:overviewInstability}).

Because we are interested in studying single-mode wave growth for these instabilities, it is convenient to use the simplifying setup of a cold, gyrotropic ring distribution of CRs, which is visualized in Figure~\ref{fig:gyro_init}, and given by
\begin{equation}
	f_\mathrm{cr} = \frac{n_\crrm}{2\pi u_\perp}\delta(u_\parallel - u_\mathrm{dr,0})\delta(u_\perp - u_{\perp,0}).
\end{equation}
Here, $\vec u=\gamma \vec \varv$ is the relativistic particle velocity, where the Lorentz factor is $\gamma=\left[1-(\vec \varv/c)^2\right]^{-1/2}$ and $\delta$ is the Dirac delta function. 
For this distribution, the pitch angle cosine $\mu = \varv_\parallel/\varv$ is fixed, while all angles $\psi$ around the parallel axis are equally likely. 
The advantage of this setup is, that it exhibits well-defined peaks in the linear dispersion relation while the physically and observationally motivated power-law distributions excite waves over a large spectrum of wave numbers $k$, making it more difficult to understand the underlying physics.

This research has been triggered after observing a strong correlation between the rotational phases of the CR velocity and the wave magnetic field in the fluid-PIC and PIC streaming simulations presented in \citet{lemmerzCouplingMultifluidDynamics2024}, which also excite a broad spectrum of waves over time. 
In those simulations, the box is large enough so that CRs cannot travel across it before the instabilities saturate, suggesting that the finite box size has no influence on the simulated instability and is rooted in plasma physical processes. 
In the following, we design a simulation suite in which we limit the simulation box size and vary the simulation parameters so that the individual CR-driven instabilities are excited separately. This helps us to analyze the instabilities and their saturation in isolation and to understand the emerging phase correlation as it is observed in our previous simulation.  

The numerical resolution samples the dispersion relation at discrete values in $k$-space \citep{shalabyImportanceResolvingSpectral2017}. 
As such, a wave mode can only be resolved in simulations with periodic boxes if the absolute value of the wave vector $\vec{k}$ is an exact multiple of $2\pi/\vec{L}$, where $\vec{L}$ is the length vector of the box.
Typically, the goal is to reproduce the analytical dispersion relation by densely sampling the modes in $k$-space, i.e., using large box sizes $L$. 
In this paper, however, we concentrate our attention on the growth of only one resonant wave vector $\vec{k}_\mathrm{res}$ and try to prevent the growth of neighboring $\vec{k}$, which would complicate the interpretation of the results because of possible mode-mode interactions masking the growth and saturation of an individual mode.
In order to achieve this, we choose the one-dimensional box length to be only a few times the scale of interest.
 That is $L_x=2\times 2\pi/k_{\mathrm{res}}$ for the \Gbsim~simulation, where $k_{\mathrm{res}}$ is the scale of the largest growing wave mode of interest, while the simulation box is $3$ times (\Gfsim) and $6$ (\Isim) times the scale $2\pi/k_\mathrm{res}$ in the other simulations.
In the \Isim simulation, this restriction also eliminates the growth of gyroscale instabilities, as Alfv\'en modes larger than the ion skin depth  $d_i = c/\omega_i$ remain unresolved.
Here, the plasma frequency for a species $s$ is given by $\omega_s = (q_s^2 n_s/m_s \epsilon_0)^{1/2}$ and the overall plasma frequency is $\omega_p=(\sum_s \omega_s^2)^{1/2}$.

On the other hand, the intermediate-scale instability is eliminated from the \Gsims~simulations by violating its growth condition, $\vdr/\valf < \sqrt{m_r}/2$ \citep{shalabyNewCosmicRaydriven2020}, where the mass ratio is given by $m_r\equiv m_i/m_e$ and the Alfv\'en velocity is $\varv_{\rm A} = B_0/(\mu_0 n_\mathrm{i} m_\mathrm{i})^{1/2}$. 
Thus, we performed this simulation with an unrealistic mass ratio of $m_r=100$ and $\vdr/\valf =10$, such that the intermediate scale would only be triggered if the particles scatter below $\vdr/\valf< \sqrt{100}/2=5 $, which is not seen in our setup.
The \Isim~simulation uses a lower $\vdr/\valf=5$ and a realistic mass ratio for two reasons: First, together with an increase of the mass ratio, this ensures that the growth condition is satisfied. 
Second, this choice moves the unstable peak of \Iesimm\ to a smaller scale of $k d_i = 362.32$, increasing the scale separation and causing it to saturate at a smaller level, as demonstrated in Sec.~\ref{sec:nonlinear_saturation}. 
The \Isim~simulation is designed to best sample the peak of the whistler regime, \Iwsimm, while suppressing the impact of \Iesimm.

As $\vdr$ is different between the \Gsims~and  \Isim~simulations, the remaining $\vperp$ parameter is chosen, such that the total velocity, $\varv=\abs{\vec\varv}\approx0.14c$, for CR ions is initially approximately the same in every simulation.
In the following, we describe the common setup for all simulations while the different parameters are given in Table~\ref{tab:simulation_parameters}.
All simulations use $75$ particles per cell for CRs per species, at a density contrast of $n_\mathrm{cr}/n_\mathrm{bg} = 10^{-4}$.
In order to enforce charge density and current neutrality, we initialize and evolve an electron beam without a perpendicular velocity but with the same drift velocity as the CR proton beam. 
The background temperature for the isotropic fluid species is set to $k_\mathrm{B} T_{s}/(m_i c^2)=10^{-4}$, where $k_\mathrm{B}$ is the Boltzmann constant and the different background species are denoted by the variable $s\in (i, e)$.
All electromagnetic fields and fluid velocities are initialized as $0$, except for the background magnetic field $B_0$, which is along the box direction, $x$.
This implies, that the background is at rest and the CRs and waves move in the simulation frame.
We set the (ion) Alfv\'en velocity $\valf = B_0/\sqrt{\mu_0 m_i n_i} = 0.01 c$. 
Note that our three simulations differ in the assumed ion-to-electron mass ratio $m_r=m_i/m_e$ and hence, also in the implicit ion cyclotron frequency of $\Omega_i = qB_0/m_i$, which serves as a physically motivated timescale.
The cell size resolves the plasma skin depth, $\Delta x = 0.1 c/\omega_p$, and the time step size resolves the speed of light $c=1$, $\Delta t=0.4\Delta x/c$.
We adopt periodic boundary conditions in our simulation domain.

We compare our results for the standard PIC and fluid-PIC methods using the parameters of simulation \Isim in Appendix~\ref{app:PIC}. This shows that the fluid-PIC method provides similar results at a significantly reduced computational cost. 

\begin{deluxetable}{lrrrrrr}%
\tablecaption{Simulation parameters including the initial CR drift and perpendicular velocities, as well as the scale and growth rate of the associated dominant resonant wave mode.}
\label{tab:simulation_parameters}
\tablehead{\colhead{Simulation} & \colhead{$L_x$}  & \colhead{$m_i/m_e$} & \colhead{$\varv_\mathrm{dr}$} & \colhead{$\varv_\perp$} &  \colhead{$k_\mathrm{res}$} & \colhead{$\Gamma_\mathrm{res}$} \\
& \colhead{$\left[c/\omega_p\right]$} & & \colhead{$\left[\valf\right]$} & \colhead{$\left[\valf\right]$} & \colhead{$\left[d_i^{-1}\right]$} & \colhead{$\left[\Omega_i\right]$}
} 
\startdata
\Gbsim & 1333.3& \phn100 & 10& 10\phd\phn & 0.095 & 0.0579\\
\Gfsim & 1735.7& \phn100 & 10 & 10\phd\phn & 0.109 & 0.0640\\
\Isim  & \phn346.8 & 1836 & 5 & 13.1 & 4.656 & 0.4880\\
\enddata
\end{deluxetable}

\section{Particle motions and wave growth}
\label{sec:theory}

\begin{figure}
   \resizebox{\hsize}{!} { \includegraphics[width=\textwidth]{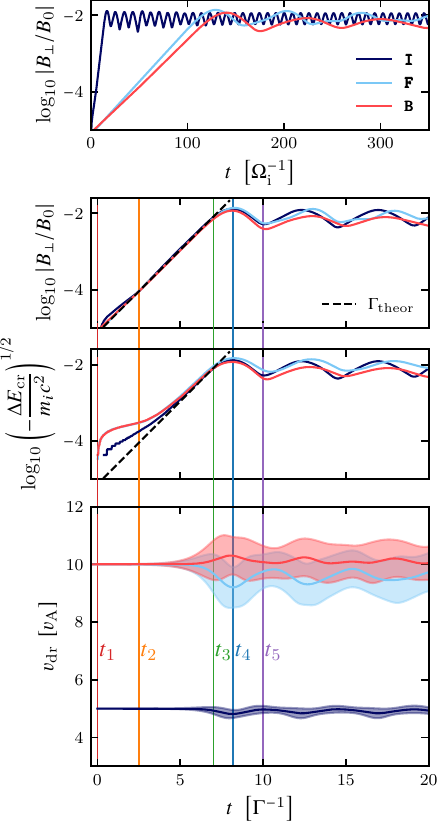} }
	\caption{Top two panels: evolution of $B_\perp$ over time for the different simulations in units of $1/\Omega_{i}$ of the corresponding simulation and in units of the respective inverse growth rate $1/\Gamma$. The latter units are useful for comparing the simulations at specific times $t_{1\textup{--}5}$. Third panel: energy lost by the CRs to the unstable modes as a function of time, which corresponds to the energy gain of the modes (see the second panel) up to numerical precision. Bottom panel: evolution of $\vdr$ over time, with one standard deviation of $\vdr$ indicating the spread around the mean value. Note that the mean velocity of CRs does not approach $\valf$ as it is usually assumed but saturates at a much larger value.}
 \label{fig:bgrowth}
\end{figure}
The interaction of the CRs with the waves can be trivially broken down into two parts: the impact of the CRs on the wave and the impact of the wave on the CRs.
In this section, we introduce the momentum equation to understand the former, as well as an evolution equation for the trajectories of individual CRs to understand the latter.
Here, we will treat the CRs in isolation without accounting for the effect of waves on the CRs and discuss the resulting shortcomings.
These equations serve then as a building block for later sections, which focus primarily on the wave-particle interaction.
\newpage

\subsection{Momentum balance}
The intensity of the growing waves is one of the most relevant quantity pertaining to CR streaming, and momentum conservation can be used to derive a useful equation relating it to changes in the CR velocity.
The CR momentum of an individual particle along the background magnetic field is $p_x = \gamma m_\crrm \varv_x$, thus the CR momentum density can be expressed as  $\mathcal{P}_\crrm = n_\crrm \bar{\gamma} m_i \vdr$, where $\bar{\gamma}$ is a relativistic prefactor obtained from averaging the CR distribution \citep{baiMagnetohydrodynamicParticleincellSimulations2019}.
Because of momentum conservation, changes in the parallel momentum density of CRs correspond to changes in the parallel momentum density of the excited electromagnetic waves.
The momentum density of the plasma waves is assumed to be stored predominantly in the movement of background particles, which needs to be taken into account. 
As the Poynting vector characterizes electromagnetic momentum without matter, which is negligible compared to the momentum carried by the background particles, it is appropriate to use the Minkowski momentum $\vec{S}_\mathrm{M}\equiv\vec{D}\cross\vec{B}$ instead, which additionally accounts for the inertia in the wave-carrying background particles. 
The electric displacement field is $\vec D=\epsilon_\mathrm{bg}\vec E$ and $\epsilon_\mathrm{bg}$ denotes the electric permittivity of the background plasma (Chapter~2 of \citealt[]{grootFoundationsElectrodynamics1972}, \citealt{kempResolutionAbrahamMinkowskiDebate2011}).
Because $\vwa=(\epsilon_\mathrm{bg} \mu_\mathrm{bg})^{-1/2}$ and the magnetic susceptibility of the background plasma is almost the same as in vacuum, $\mu_\mathrm{bg}\approx\mu_0$, the parallel momentum of the wave is 
\begin{equation}
S_\mathrm{M}=\left.\vec E \cross \vec B/(\mu_0 \vwa^2)\right|_\parallel = B_\perp^2/(\mu_0 \vwa).
\end{equation}
The last equality assumes a single, transversal wave mode traveling at a phase speed of $\vwa$, for which $\vec{E}_\perp = -\ci\vwa \vec{B}_\perp$ follows according to Faraday's law.
Evaluating the momentum balance of CR momentum lost by driving the unstable wave yields
\begin{equation}
\Delta S_\mathrm{M} + \Delta \mathcal{P}_\crrm =0  ~~ \Rightarrow ~~
 \frac{\Delta B_\perp^2}{B_0^2} = - \frac{n_\crrm}{n_\mathrm{bg}} \frac{\vwa \Delta \left(\bar{\gamma}  \vdr\right)}{\valf^2}, \label{eq:CRmomentumeq}
\end{equation}
where $\Delta x=x(b)-x(a)$ is the difference between the times $a$ and $b$.\footnote{For forward moving waves ($\vwa>0$), CRs slow down in the linear regime so that $\Delta \vdr<0$.}
This means, that the wave intensity mostly depends on the difference in drift velocity of the CR population. 

The magnetic field growth for both simulations is shown in  Figure~\ref{fig:bgrowth}. 
We note, that the saturation levels of the \Gsims~and \Isim~simulations do not necessarily coincide if we were to use the same initial CR pitch angle. We postpone a systematic study of this topic to future work.

Although instability growth at the gyroscale takes significantly longer in physical time units, all simulations exhibit a similar behavior when the time is scaled to their maximum growth rates $\Gamma$.
We mark $5$ times of interest: first, the initialization, second, the phase of linear growth, third, the transition to the nonlinear regime, fourth, the time of saturation, and fifth, the rebound point, to which we will refer throughout the paper.
The momentum equation~\eqref{eq:CRmomentumeq} states that changes in the drift velocity affect the magnetic field strength as $\Delta B_\perp^2 \propto -\vwa \Delta\vdr$. 
The simulation~\Gbsim~is qualitatively different from the other simulations as the CRs are accelerated rather than slowed down in the parallel direction. 
This is expected from the momentum equation because $\vwa$ is negative and as energy is transferred to these backward-propagating modes, $\vdr$ needs to increase over time.
However, we can infer from the third panel of Figure~\ref{fig:bgrowth} that CRs still lose energy, which stems from a decrease in perpendicular velocity $\varv_\perp$.

The wave velocity of the fastest driven modes of the intermediate-scale instability is faster than that driven by the gyroscale instabilities, i.e., $\varv_{\mathrm{wave},\Isimm} \approx 6.52 \varv_{\mathrm{wave},\Gfsimm} $. Thus, it generates a larger magnetic field with the same $\Delta \vdr$ as can be inferred from equation~\eqref{eq:CRmomentumeq}. 
As a result, the pitch angle scattering of the intermediate-scale instability is significantly reduced because similar levels of magnetic field amplification are reached in all simulations.
This effect is captured in the standard deviation around the drift velocity, which serves as a measure of this pitch angle scattering and can be compared between the simulations (see the bottom panel of Figure~\ref{fig:bgrowth}).

After saturation, all simulations show oscillatory periods of wave  growth and decay (corresponding to particle acceleration and deceleration) with a similar periodicity. This oscillatory behavior in the wave intensity is observed in most single wave mode instabilities, e.g., the electrostatic two-beam instability \citep{morseOneTwoThreeDimensional1969, shoucriNonlinearEvolutionBump1979}, and beam-plasma instabilities \citep{Shalaby+2018,Shalaby+2020}.

\begin{figure*}
   \resizebox{\hsize}{!} { \includegraphics[width=\textwidth]{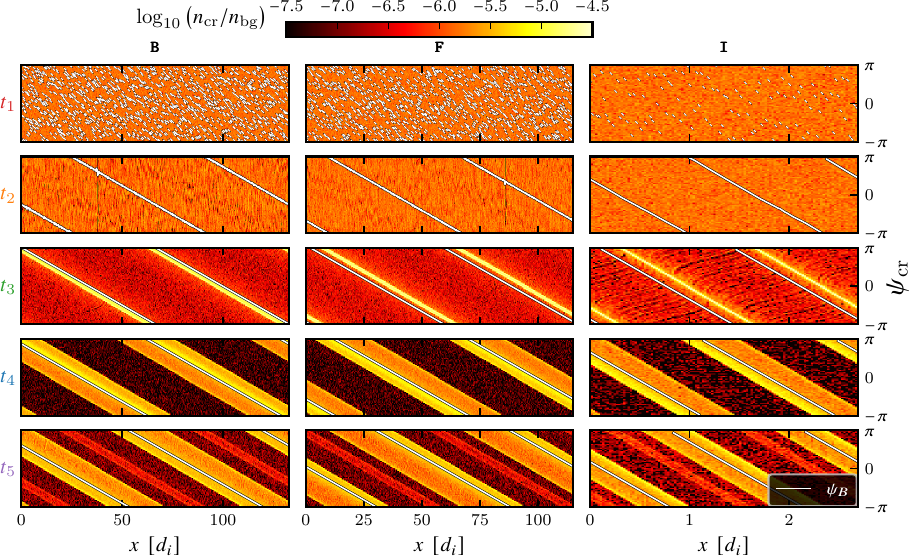} }
	\caption{We show the CR distribution as a function of local rotational phase of CR ions, $\psi_\crrm$, and the position along the initial magnetic field, $x$, at the five different times $t_{1\textup{--}5}$ as defined in  Figure~\ref{fig:bgrowth}.
	These are overplotted with white lines, indicating the local rotation phase of the perturbed magnetic field vector, $\psi_B$. 
	The angle $\psi_B$ of the magnetic wave follows a straight, white line with slope of $-k$, according to $\psi_B \propto \arg\{B_\perp \exp[\ci\,(\omega t_i - k x)]\}=-k x$, except at initialization where it is random.
	An illustration of the particle resonance is shown in Figure~\ref{fig:resonance}.
	The average CR density is $n_\crrm(x)=10^{-4} n_\mathrm{bg}$,  which is retrieved when contracting the $\psi$ dimension in this plot. We only show a part of the simulation boxes so that $2$ cycles of the dominant wave mode are captured in all plots. Clearly, the action of the instability causes the CR phases to bunch up close to the local phase of the excited magnetic field.}
	\label{fig:xtscatter}
\end{figure*}

\begin{figure*}
	\plotone{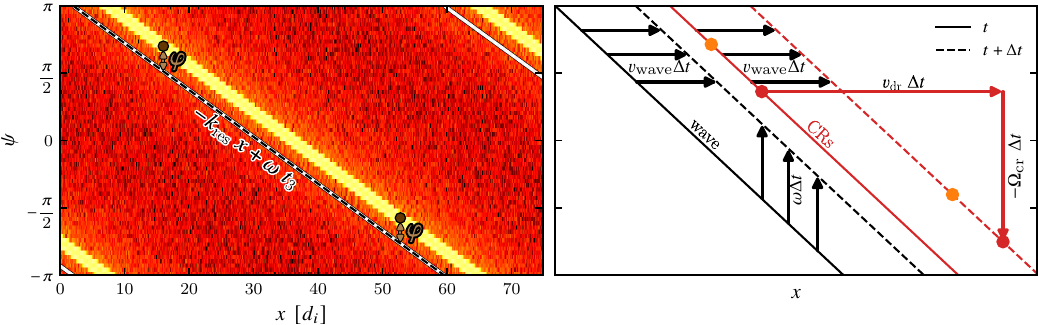}
	\caption{The left panel shows a zoom into the middle panel (\Gfsim~at $t_3$) of  Figure~\ref{fig:xtscatter}. The electromagnetic wave  is shown to have a slope of $-k_\mathrm{res} x$, while two particles are indicated at a relative angle $\varphi$.
	The right panel shows an illustration of the left panel, which explains the conundrum that the CRs follow a helical structure that has the same winding angle as the magnetic field of the unstable wave, despite the fact that they are much faster than the wave. We show the wave and the dense, narrow CR band while we omit their periodic wrapping for clarity. According to the resonance condition, a CR particle (red) not only moves at $\vdr$ but also rotates at its gyrofrequency $\Omega_\crrm$. Thus, the particles distributed along the helix at $t$ (solid red) are mapped onto a somewhat displaced helix at $t+\Delta t$ (dashed red). By contrast, the wave moves by $\varv_\mathrm{wave} \Delta t$ along $x$, but its movement can alternatively be understood as a rotation of $\omega \Delta t$ along $\psi$, thus mapping the wave from $t$ (solid black) to $t+\Delta t$ (dashed black). This explains how the CR helix maintains the same distance from that of the magnetic field vector of the wave.
	}
	\label{fig:resonance}
\end{figure*}

\begin{figure}
   \resizebox{\hsize}{!} { \includegraphics[width=\textwidth]{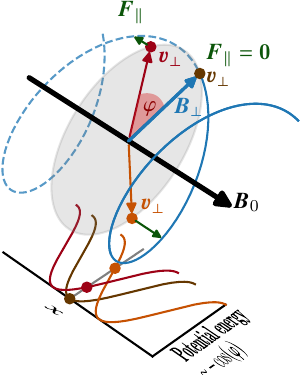} }
	\caption{We show three CR particles at the same position $x$ and perpendicular velocity $\varv_\perp$, which are in resonance with an electromagnetic wave (shown in blue), i.e., the angle $\varphi$ changes only slowly over time.
	While the particles share a spatial position $x$, they experience different parallel forces, which depend on their gyrophases $\varphi$.
	The Lorentz forces on the particles act to minimize $\varphi$, meaning that the particles align themselves with the local orientation of the magnetic field vector. 
	This is achieved by means of accelerating (decelerating) the particle in the parallel direction, which in turn increases (decreases) the Doppler-shifted gyration of the particles, $k \varv_x - \Omega_\crrm$, and thus acts as a pseudo-force in the $\varphi$ direction.
	The particle's trajectory can be shown to follow a pendulum motion in a potential well that is centered on the local orientation of the perpendicular magnetic field.
	}
	\label{fig:pendulum_potential}
\end{figure}

\subsection{Evolution of the instability without CR back-reaction: the pendulum equation}

While the evolution of $\vdr$ plays a crucial role, it is instructive to study the angle $\psi_\crrm=\arg(\vec\varv_\perp)$ of the particles (cf.\  Figure~\ref{fig:gyro_init}).
For all perpendicular vectors, we use the shorthand complex notation $\vec\varv_\perp = (\varv_y + \ci\,\varv_z) \vec{e}_\perp$, where $y$ and $z$ span the plane perpendicular to $\vec{B}_0$, see also Appendix~\ref{app:convention} for our notation convention.  
It simply follows, that $\varv_\perp=\abs{\vec\varv_\perp}$.

To motivate the following calculations, we first analyze the simulated structure of the distribution of rotation angles of CR ions, $\psi_\crrm$, as well as $\psi_B=\arg{\vec B_\perp}$.
These are shown as CR distributions and lines, respectively, as a function of $x$ position and rotation angles in Figure~\ref{fig:xtscatter}. 
The magnetic field is initially randomly aligned but the dominant wave mode is quickly excited and structures the perpendicular magnetic perturbation at $t_2$.  
The CR distribution is still mostly uniform, but changes significantly before entering the nonlinear stage at $t_3$. 
At every position $x$, the CRs have now bunched up to a narrow distribution in $\psi_\crrm$ so that we obtain a broader helical structure that winds around the mean magnetic field $B_0$. The CR helix has exactly the same winding angle in comparison to the helix delineated by the unstable magnetic wave, but the helical structures of the CRs and the magnetic wave are offset with respect to one another: we obtain $\psi_\crrm>\psi_B$ for forward moving waves and $\psi_\crrm<\psi_B$ for backward moving waves.
At saturation, the spread in the $\psi_\crrm$ angles is again larger so that they form a broader strip that extends over the magnetic field line.
At $t_5$ a \enquote{ghost} strip can be seen, which is the result of particles escaping from the main strip to the left and to the right, overlapping in between.
Even though the spatial and temporal scales are very different, the \Gfsim~and \Isim~simulations share the same features.

When comparing the distributions of CRs and magnetic perturbations in Figure~\ref{fig:xtscatter}, it is obvious that $\psi_\crrm$ and $\psi_B$ are closely related.
It is therefore useful to define the gyrophase 
\begin{equation}
	\varphi(x,t) = \psi_\crrm(t) - \psi_B (x,t)=\arg(\vec \varv_\perp \vec B_\perp^\dagger)(x,t) 
	\label{eq:phi_definition}
\end{equation} 
for each particle.
Here, $\dagger$ denotes the complex conjugate.
Essentially, the particle angle $\varphi$ is now defined in a helical coordinate system, where the helix is given by the electromagnetic wave.
We can estimate the gyration period as $\psi_\crrm(t)\sim-\Omega_\crrm t$ (where $\Omega_\crrm=\Omega_i/\gamma$) while the moving particles experience the magnetic field at $\vec B(x_0+\vdr t , t)$. 
Given that $\arg(\vec B_\perp)(x, t) = - k x + \omega t$ (for parallel waves with phase speed $\vwa$ and rotation rate $\omega=k \vwa$), it follows trivially that the gyrophase changes over time as
\begin{eqnarray}
    \varphi(x,t) &=& k (x_0 + \vdr t) - \omega t - \Omega_\crrm t 
    \nonumber\\
    &=&\varphi_0 + [k (\vdr  -  \vwa) - \Omega_\crrm] t,
	\label{eq:phi_simplified}
\end{eqnarray}
where we chose $\varphi_0=k x_0$. Enforcing $\varphi(x,t)$ to be approximately constant over time, we recover the resonance condition \citep{kulsrudPlasmaPhysicsAstrophysics2004} 
\begin{align}
     \mathcal{R}(\omega, k) &\equiv 
     k (\vdr  - \vwa) - \Omega_\crrm \nonumber \\
     &= k\phantom{(} \vdr - \mathmakebox[\widthof{\vwa)}][l]{\omega} - \Omega_\crrm
     = 0.
     \label{eq:resonance}
\end{align}
From this condition, one can find multiple waves with a given $\omega(k)$ and $k$, which are resonant.
Furthermore, equation~\eqref{eq:resonance} implies that resonant particles move in lockstep along the wave. 
This picture is geometrically illustrated in Figure~\ref{fig:resonance}.
Interestingly, even though single particles move significantly faster than the wave, collectively they experience the wave as a static electromagnetic field.
Thus, the CRs form a coherent, wave-like structure, which moves at velocity $\vdr - \Omega_\crrm/k$. 
This excites waves, which move at the same velocity.
Furthermore, if $\vdr < \Omega_\crrm/k$, then the CR band moves backwards and can excite backward moving waves.

We can expand on the previously described simplified picture, which only included the force exerted by $B_0$ on a particle.
In the following, we compute a single particle trajectory that results from the full Lorentz force introduced by the wave.
In the wave frame, we can neglect contributions to the electric field so that the CR particle energy and its relativistic Lorentz factor $\gamma'$ remain constant. For this reason, we adopt this frame in order to derive the time evolution of the particle velocity with $\vparwave = \varv_\parallel - \vwa$ and $\varv_\perp'=\varv_\perp$:
 \begin{equation}
	\pdv{[\gamma'\vparwave (t)]}{t} = \frac{q}{m} {\left[\vec{\varv}' \cross \vec{B}\right]}_\parallel = -\frac{q}{m} B_\perp (t) \varv_\perp'(t)\sin \left( \varphi(t) \right).
	\label{eq:lorentz_varx}
\end{equation}

An equation for the perpendicular particle velocity can be easily derived by using the energy conservation in the wave frame, that is $\partial_t (\varv_\perp'^2+\vparwave^2)=0$.
Thus,
\begin{equation}
	\label{eq:vperpdt}
	\pdv{[\gamma'\varv_\perp'(t)]}{t} = -\frac{\gamma'\vparwave}{\varv_\perp'} \pdv{\vparwave}{t} = \frac{q}{m} B_\perp (t) \vparwave(t)\sin \left(\varphi(t) \right),
\end{equation}
which is a projection of the Lorentz force term $\vec{\varv}_\parallel' \cross \vec B_\perp q/m$. Specifically, if $\vec\varv_\perp$ points along this Lorentz force term (i.e., for $\varphi=\pi/2$) only the magnitude of $\vec\varv_\perp$ is increased without changing its direction. In the case of $\vec\varv_\perp\parallel \vec B_\perp$ (i.e., for $\varphi=0$), the length of $\vec\varv_\perp$ remains invariant, but the Lorentz force on the particle causes it to change its rotational velocity. 
Assuming that both $\psi_\crrm$ and $\psi_B$ are measured from the same starting point in the plane perpendicular to $\vec{B}_0$, the remaining part of the Lorentz force term $\vec{\varv}_\parallel' \cross \vec B_\perp q/m$ is projected onto $\psi$ such that the angular velocity of the particle is
\begin{equation}
	\pdv{\psi_\crrm}{t} = - \Omega_\crrm + \frac{q}{\gamma' m}\vparwave B_\perp (t) \frac{\cos(\varphi(t))}{\varv_\perp (t)}.
	\label{eq:lorentz_psi}
\end{equation}

In equation~\eqref{eq:phi_simplified}, we assumed $ \varv_\parallel = \vdr$ at all times. However, the $x$-coordinate of an individual particle is correctly defined as $x(t) = \int_{0}^{t} \varv_\parallel(\tau) \mathrm{d}\tau$.
Taking the time derivative of $\varphi$ (as defined in equation~\ref{eq:phi_definition}) and eliminating $x(t)$ and $\partial_t\psi_\crrm$ (equation~\ref{eq:lorentz_psi}) yields 
\begin{equation}
\pdv{\varphi}{t} = -\Omega_\crrm + k \vparwave(t) + \frac{q}{\gamma' m}\vparwave(t) B_\perp (t) \frac{\cos(\varphi(t))}{\varv_\perp' (t)}.
	\label{eq:phi_full0}
\end{equation}
Due to the different time dependent quantities, this equation is complicated to solve.
As before, $\partial{\varphi}/\partial t = 0$ can be interpreted as a resonance condition.  However, we further make the approximation that $q B_\perp /(\gamma' m \varv_\perp) \ll k$, which is equivalent to $B_\perp/B_0 \ll k d_i \times \varv_\perp/\valf$, and drop the last term of equation~\eqref{eq:phi_full0} from  subsequent calculations.  
Thus, we retrieve the same resonance condition as before, but now in the comoving wave frame.
With this simplification, the evolution of $\varphi$ can be further investigated, yielding
\begin{align}
	\varphi(t) &= -\Omega_\crrm t + k \int_{0}^{t} \varv_\parallel'(\tau) \mathrm{d}\tau +\varphi_0,\\
	\pdv{\varphi (t)}{t} &= -\Omega_\crrm  + k  \varv_\parallel'(t),\label{eq:CRdphidt}\\
	\pdv[2]{\varphi (t)}{t} &= -k \frac{q B_\perp (t) \varv_\perp(t)}{\gamma' m} \sin \varphi (t) \label{eq:CRpendulum}.
\end{align}
This shows that in the limit of weak perturbations, the angle $\varphi$ between $\vec\varv_\perp$ and $\vec B_\perp$ obeys a pendulum equation. We also see that the parallel Lorentz force from equation~\eqref{eq:lorentz_varx} multiplied with the wavenumber $k$ can be interpreted as a pseudo torque on $\varphi$. This is because the gyrophase is defined in relation to a helical coordinate system spanned by $\vec B_\perp$. This pseudo torque is absent from the evolution of $\psi_\crrm$, which is defined in an inertial frame. 

While Equation \eqref{eq:CRdphidt} shows that resonantly driven waves will not change the relative phase between the CR and the local wave magnetic field (i.e., up to zeroth order in $B_\perp/B_0$) if the parallel velocity stays constant. Up to first order, these waves exert a parallel force given in Equation \eqref{eq:lorentz_varx} such that the particles accelerate towards locations where $\varphi(t)$ is close to zero.

\subsection{Discussing the pendulum picture of CR motions}

Figure~\ref{fig:pendulum_potential} visualizes the parallel Lorentz force acting on three test particles at the same position $x$ in the gray plane.
Particles aligned with the perpendicular magnetic field do not experience a parallel Lorentz force while the unaligned particles are moving parallel to $B_0$ towards the closest field line.
As they move along $x$, their relative gyrophase $\varphi=\psi_\crrm-\psi_B$ is minimized -- while this movement along $x$ has no influence on the evolution of the angle $\psi_\crrm$ (equation~\ref{eq:lorentz_psi}) in the static coordinate system.

Interestingly, equation~\eqref{eq:CRpendulum} is equivalent to the pendulum differential equations, and thus, the CRs are trapped in potential wells that are centered on the local direction of the magnetic perturbation, around which they oscillate. 
This potential well not only depends on the local magnetic field strength $B_\perp$, but also on the gyrophase of each particle. 
This differentiates it from a magnetic bottle, which is localized in space.
Note that conservation of the adiabatic moment $\mu=\gamma m \varv_\perp^2/(2 B)$ cannot be used to understand resonant CRs because there is no effective cyclotron motion with respect to the electromagnetic wave, which precludes the applicability of the adiabatic assumption.

If the wave amplitude saturated its growth and if changes in the CR velocities are small, the change in the pitch angle of a CR due to an interaction with a wave packet can be approximated \citep[Chapter~12.2.\ of][]{kulsrudPlasmaPhysicsAstrophysics2004}.
Starting from a gyrotropic distribution of CRs, we are faced with another problem: the time average of the term $\int_{-\pi}^{\pi} \sin(\varphi(t)) \mathrm{d}\varphi \sim0$ averages out, which is equivalent to stating that $\dot{\varphi}$ averages out for the CR population. This in turn means that $\Delta \vdr(t)\sim0$. Thus, no wave growth would occur according to the momentum equation~\eqref{eq:CRmomentumeq}.

This is contrary to what our simulations and solutions of the dispersion relation show. 
One possible approach for explaining the surplus in transferred momentum assumes the interaction with individual wave packets of length $d$. 
The particles transit through the wave packet in time $t_d = d/\varv_\parallel$. Hence, particles that are accelerated in $\varv_\parallel$ traverse the wave packet faster than decelerated particles.  
The change in momentum depends on the force times the time spent interacting with the wave packet, $\Delta p_\parallel = F_\parallel t_d$. Because $t_d$ is smaller for fast particles, a stochastic imbalance between accelerated and decelerated particles occurs. 
If, for a forward moving wave, the faster CRs take momentum from the wave and the slower CRs give momentum to the wave, there would be a surplus of momentum given to the wave because of the longer interaction times of slower particles, which would amplify the wave.
This mechanism is similar to second order Fermi acceleration, however particles are not reflected but pass through the wave packet and thus lose energy \citep{fermiOriginCosmicRadiation1949,tsytovichMechanismWaveAbsorption1985}.

However, this argument would also predict a damping of backward moving waves, for which slower CRs take momentum from the wave and faster CRs give momentum to the wave, even though the unstable waves are still expected to grow. 
Because our simulations use a periodic box, $d$ is effectively infinitely long and this effect is eliminated in our setup. 
Therefore, instability growth cannot be caused by differences in the transit time and motivates the search for another explanation.
In Section~\ref{sec:linear_growth} we investigate the underlying mechanism leading to a surplus of CRs giving momentum to the wave, creating the imbalance that is necessary for wave growth.

We point out two more intricacies, which differentiate the description of parallel CR motions from a traditional pendulum.
A traditional pendulum oscillates at a frequency of $(g/l)^{1/2}$, where $l$ is the length of the pendulum and $g$ is the gravitational acceleration, both of which are approximately constant.
Analogously, the CR pendulum frequency depends on the amplitude of $B_\perp(t)$.
During wave growth, $B_\perp(t)$ grows exponentially and is even closely related to $\Delta \varv_\parallel$ through the momentum equation~\eqref{eq:CRmomentumeq}.
Thus, the gyrophase of CRs in the linear growth phase resembles a pendulum whose length is shortened exponentially over time.
Second, these equations are derived for the interaction of a single CR with a single wave mode and constant $\Omega_\crrm$, while a realistic situation has many CRs interacting with multiple wave modes.
A traditional analogue is a coupled pendulum, which further complicates an accurate analytical treatment.

\begin{figure*}
   \resizebox{\hsize}{!} { \includegraphics[width=\textwidth]{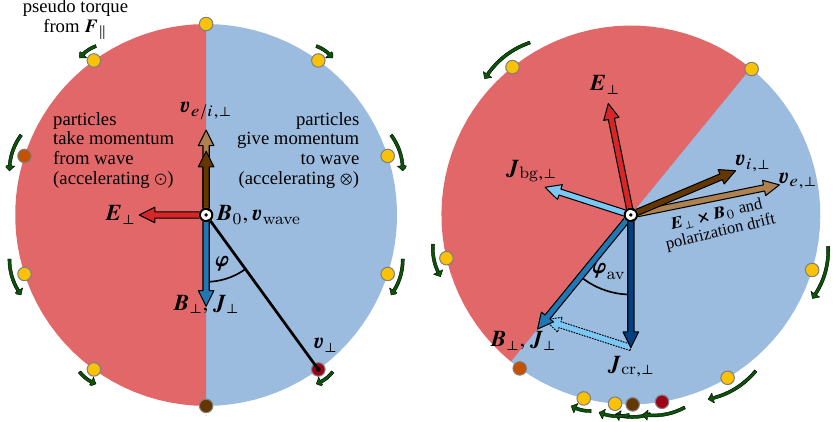} }
	\caption{A drawing explaining the processes leading to instability. We show the $\varphi$ angle of the streaming CRs in relation to their local direction of the perpendicular magnetic and electric fields, $\vec{B}_\perp$ and $\vec{E}_\perp$, respectively. On the left-hand side, we show the initial seed wave with random particles at a given position $x$ and use the same color coding of the three particles defined in Figure~\ref{fig:pendulum_potential} (while we color code the additional particles with yellow). As explained in  Figure~\ref{fig:pendulum_potential}, the parallel Lorentz force, $q \vec \varv_\perp \cross \vec B_\perp$, causes all particles except for the brown one to move along $\vec{B}_0$ and $\vec{x}$, implying a change in $\varphi(t)$, which we indicate by green arrows at each particle. Particles on the left (in the red semicircle) get accelerated out of the plane into the propagation direction of the wave, and hence take momentum from it to ensure momentum conservation while particles on the right (blue semicircle) are accelerated into the plane and transfer momentum to the wave. Thus, we would expect all particles to converge to the local magnetic field direction $\vec{B}_\perp$ (albeit at different locations on the $x$ axis). This situation is shown in the drawing on the right-hand side, which shows the particle distribution in $\varphi$ at a later time. However, this bunching up of CRs implies a downwards pointing CR current, $\vec{J}_{\crrm,\perp}$. After adding the background current, $\vec{J}_{\mathrm{bg},\perp}$, the total current, and hence the perturbed magnetic field $\vec{B}_\perp$, is shifted to the left. Thus, the back-reaction of the CR current causes more CRs to be found in the blue region, in which there is a net transfer of CR momentum to the wave. This explains the inner workings of the resonant instability. The background velocities $\vec{\varv}_{i,\perp}$ and $\vec{\varv}_{e,\perp}$ determine the orientation of $\vec{J}_{\mathrm{bg},\perp}=n_\mathrm{bg}(\vec{\varv}_{i,\perp} - \vec{\varv}_{e,\perp})$ while $\varphi_\mathrm{av}$ denotes the averaged phase angle of the CRs.}
	\label{fig:pendulum_schematic}
\end{figure*}

\section{The physics of wave growth and decay}
\label{sec:linear_growth}

As derived earlier, the gyro angles of CRs obey the pendulum equation with the force pointing in the direction of the local $\vec{B}_\perp$. 
However, we do not observe the oscillating behavior of a pendulum during the linear growth phase but only after saturation.  
In this section, we concentrate on the linear phase and the physical mechanism behind wave growth.
This section is structured in the following way:
first, we provide an intuitive physical picture of the resonant wave growth and discuss its implications.
To this end, the different equations and corresponding effects are discussed in succession.
This model is then compared with the dispersion relation, which is an exact solution to the linearized wave equation and the CR Vlasov equation, capturing all effects simultaneously while making it difficult to extract a simple physical meaning underlying the equations.
Finally, we connect these considerations through our simulation results. 

\subsection{Deconstructing the instability's feedback loop}

Exponential growth processes often have an underlying feedback loop, which we will describe for the resonant CR-driven instabilities in the following.
In essence, the CRs try to align their gyrophase with the perpendicular magnetic field, as pointed out in  Figure~\ref{fig:pendulum_potential}. However, the resulting CR current does not only intensify the wave, but also modify its wave speed.
Thus, the wave and the associated potential wells move constantly, but slowly away from the particles at resonance -- leading to an asymmetry, which on average forces CRs to transfer momentum to the wave. 
In the following, we detail the individual physical processes leading to instability growth for a forward moving wave (defined by $\vwa>0$) with a wave vector $k>0$ (cf.\ Figure~\ref{fig:pendulum_schematic}).\footnote{In our convention, we have $\vec{a}_\parallel \cross \vec{b}_\perp = \ci\, a_\parallel \vec{b}_\perp$, $\vec{\nabla} \rightarrow-\ci\, k$, $\partial_t  \rightarrow (\ci\, \omega+\Gamma)$ so that $\vec{\nabla}\cross\vec{b}_\perp = -\ci^2\, \vec{b}_\perp$ (Appendix~\ref{app:convention}). A phase shift by $\ci$ corresponds to a 90$^\circ$ counterclockwise rotation in Fig.~\ref{fig:pendulum_schematic}.\label{footnote1}}
\begin{itemize}
	\item A (seed) electromagnetic wave travelling at $\vwa$ introduces an electromagnetic field perpendicular to its propagation direction. This is the starting point of the initial magnetic bunching provided by the magnetic perturbation $\vec{B}_\perp$ of the seed wave.
	\item CRs are accelerated by the parallel Lorentz force, and hence experience a pseudo-torque by moving along the propagation direction of the rotating wave (cf.\ Figure~\ref{fig:pendulum_potential}). In result, the parallel motion of the CRs decreases $\varphi$ and thus accelerates them toward $\vec{B}_\perp$ with an amplitude depending on $\vec{B}_\perp$ (magnetic bunching).	
	\item As the CRs' perpendicular velocity vectors are bunching up in $\varphi$, this generates a perpendicular CR current density $\vec{J}_{\crrm, \perp}$ (see Figure~\ref{fig:pendulum_schematic}).
    \item The seed electric field $\vec{E}_{\mathrm{wave},\perp}$ is perpendicular to the magnetic field. Adding the CR current, $\vec{J}_{\crrm, \perp}$, induces an additional electric field, $\vec{E}_{\crrm,\perp}$, which opposes this current. As a result, the total electric field, $\vec{E}_\perp=\vec{E}_{\crrm,\perp}+\vec{E}_{\mathrm{wave},\perp}$ is no longer perpendicular to $\vec{B}_\perp$.
    \item The electric field $\vec{E}_\perp$ leads to guiding center drifts of the background species $s$. This is the dominant effect for the background species because their perpendicular and drift velocities are small and as such, their magnetic bunching due to the term $\vec \varv_{s,\perp} \cross \vec B_\perp$ is negligible. These guiding center drifts are the $\vec{E}\cross\vec{B}$ drift \citep{Sturrock1994},
    \begin{equation}
	   \vec{\varv}_{E\cross B} = \vec{E}_\perp\cross \vec{B}_0/B_0^2  = -\ci \vec{E}_\perp / B_0,
	 \end{equation} 
  and the polarization drift \citep{Sturrock1994},
	 \begin{equation}
	  \vec{\varv}_{\mathrm{pol}}= (\partial \vec E/\partial{t}) /(\Omega_s B_0) =  (\ci\, \omega + \Gamma) \vec E_\perp /(\Omega_s B_0),
	 \end{equation}
  where we assumed a single transverse, plane-parallel wave in the last step.\footnote{The drift for Alfv\'en waves is dominated by the polarization drift of the ions as the $\vec{E}\cross\vec{B}$ currents of ions and electrons exactly cancel each other. For whistlers, the guiding center approximation of the ions breaks down so that they can be considered to be immobile on this scale, leaving the electron $\vec{E}\cross\vec{B}$ drift.} Combining both drifts introduces a perpendicular velocity component for the background ions and electrons:
   \begin{equation}
       \vec{\varv}_\perp = \vec{\varv}_{E\cross B} + \vec{\varv}_{\mathrm{pol}}.
   \end{equation}
	This leads to a perpendicular current from the background ions and electrons, 
	\begin{equation}
		\vec{J}_{\mathrm{bg},\perp} =  n_\mathrm{bg}( q_i \vec \varv_{i,\perp} + q_e \vec \varv_{e,\perp})=n_\mathrm{bg}(\vec{\varv}_{i,\perp}-\vec{\varv}_{e,\perp}).
	\end{equation}
	\item The induced magnetic field is well approximated using Amp\`ere's law without the displacement current (see footnote~\ref{footnote1}), 
	\begin{equation}
		\vec \nabla \cross \vec B_\perp= k_\parallel \vec{B}_\perp = \mu_0 \vec{J}_\perp = \mu_0 (\vec{J}_{\crrm,\perp} +  \vec{J}_{\mathrm{bg},\perp}).	
  \label{eq:J_induction}
  \end{equation}  
According to this equation, $\vec{J}_\perp$ and $\vec{B}_\perp$ are necessarily aligned for $k>0$. Imagine a situation, where an initial $\vec{B}_\perp$ gives rise to the magnetic bunching of CRs and the background particle drifts described above. The resulting total perpendicular current is not aligned with this initial $\vec{B}_\perp$. But the induced change of the magnetic field by the total current will realign $\vec{B}_\perp$ with $\vec{J}_\perp$ and hence rotates $\vec{B}_\perp$ in the perpendicular plane.
  \item  Because the magnetically bunched CR current $\vec{J}_{\crrm,\perp}$ is misaligned with $\vec{J}_{\mathrm{bg},\perp}$, so are $\vec{J}_{\crrm,\perp}$ and $\vec{B}_\perp$.
	 As a result, the average Lorentz force on the CRs, $\vec J_{\crrm,\perp} \cross \vec B_\perp$ leads to a parallel deceleration of the CRs on average (see Figure~\ref{fig:pendulum_schematic}).  
 This parallel momentum is transferred from the CRs to the background particles and the corresponding wave, as $\vec{J}_{\mathrm{bg},\perp} \cross\vec B_\perp = - \vec{J}_{\mathrm{cr},\perp} \cross\vec B_\perp$ (which is obtained by taking the cross product of equation~\eqref{eq:J_induction} with $\vec B_\perp$). 
   These changes in momentum are directly coupled to the wave intensity, according to the momentum equation~\eqref{eq:CRmomentumeq}, and thus lead to wave growth. \\
    \item The graphical representation of this feedback loop in Figure~\ref{fig:pendulum_schematic} reveals that as the CRs try to align their perpendicular velocities with $\vec{B}_\perp$, $\vec{B}_\perp$ rotates away from them.
    CRs try to realign with the rotated $\vec{B}_\perp$, which thus leads to a constant rotation of $\vec{B}_\perp$.
 	   This is best described as a frequency shift of the wave rotation rate by $\omega\rightarrow\omega+\delta \omega$, which represents a rotation in the resonance condition  
     \begin{equation}
 	\mathcal{R}(\omega+\delta \omega, k)=-\Omega_\crrm + k \vdr - ( \omega + \delta\omega)= -\delta\omega,
     \end{equation}
      as opposed to the expected resonance condition $\mathcal{R}(\omega, k)=0$ (see equation~\ref{eq:resonance}). For forward moving waves, we have $\delta\omega<0$, which implies that the wave frequency is reduced and $\omega+\delta\omega<\omega$.
\end{itemize} 

In Figure~\ref{fig:pendulum_schematic}, we assumed that $\vwa$ is aligned with the direction of $B_0$. 
Wave growth only occurs at $k>0$ (according to our convention), but growth is not constrained by the direction of $\vwa$, which can be either positive or negative.
Growth of backwards moving Alfv\'en waves is explained by simply mirroring the field vectors of Figure~\ref{fig:pendulum_schematic} about its vertical axis.
When changing the sign of $\vwa$, the sign of $\omega$ changes likewise.
The direction of $\vec{J}_{\mathrm{bg}}$, which stems from the polarization drift (proportional to $\omega$), thus changes sign as well.  
The drawing on the right-hand side of Figure~\ref{fig:pendulum_schematic} would then show the magnetic field $\vec B_\perp$ and $\vec J_\mathrm{bg}$ preceding $\vec{J}_\crrm$. 
This would lead to a parallel acceleration of CRs on average and (as the sign of the momentum equation switches likewise) wave growth of the backward moving wave. 
Similar to the forward moving wave, the wave rotation is counteracted as well, that is $\delta \omega>0$. 
The growth of a backward moving wave can be observed in  Figure~\ref{fig:bgrowth}, which shows that the particle drift velocity is growing over time, while the $\varphi$ angle during growth is on average less than $0$, as shown at $t_3$ in the left panels of  Figure~\ref{fig:xtscatter}. 
In the latter figure, it is instructive to directly compare the backwards moving wave of \Gbsim~with the forward moving wave of \Gfsim, revealing the mirroring of the particles with respect to the field vector.

A corollary of these considerations is that the induced wave velocity, $\varv_\mathrm{ind}$, is always slower than the pristine wave velocity without CRs, $\varv_\mathrm{prist}$, irrespective of whether it propagates in the forward or backward direction. This can be seen by considering forward moving waves ($\omega>0$ and $\delta\omega<0$), which obey $\varv_\mathrm{ind}=(\omega+\delta\omega)/k<\omega/k=\varv_\mathrm{prist}$. For backward moving waves ($\omega<0$ and $\delta\omega>0$), we also have a slower moving induced wave as $\varv_\mathrm{ind}=|\omega+\delta\omega|/k<|\omega/k|=\varv_\mathrm{prist}$.

Wave damping, on the other hand, can occur in two ways in the picture presented here. First, if the particle rotation overtakes the perturbed wave rotation, and second, for wave modes at negative values of $k$. We will focus on the latter effect in this paragraph, while the former is discussed in Section~\ref{sec:nonlinear_saturation}.
If there were a magnetic wave with $k<0$, the CRs attempt to bunch up towards the perpendicular wave magnetic field at some initial time, $B_{\mathrm{init},\perp}$. However, this bunching CR current will induce a magnetic field,  $k_\parallel \vec{B}_{\mathrm{ind},\perp} = \mu_0 \vec{J}_{\crrm,\perp}$, which is oriented opposite to $B_{\mathrm{init},\perp}$, thereby reducing the wave amplitude to approximately $B_{\mathrm{init},\perp}-B_{\mathrm{ind},\perp}$. Because the bunching efficiency depends on this field amplitude (which decreases over time), this describes a negative feedback loop and implies wave damping. Thus, only waves with $k>0$ (according to the convention used here) can initially grow in our simulation, which includes forward and backward traveling waves ($\vwa>0$ and $\vwa<0$).

\begin{figure*}
   \resizebox{\hsize}{!} { \includegraphics[width=\textwidth]{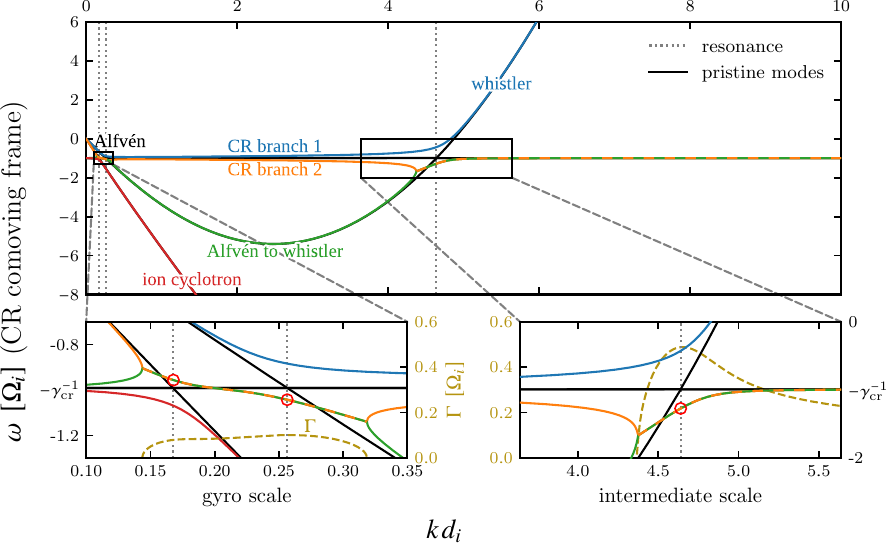} }
	\caption{The solution of the dispersion relation~\eqref{eq:fulldisp-CR} in the CR frame (see equation~\ref{eq:resonance_CR}) using the parameters of the \Isim~simulation. Solid colored lines show the real frequency, while dashed colored lines show the growth rate of the wave modes. 
    Solid black lines indicate the dispersion relation of the pristine background modes and CR ion-cyclotron waves without taking into account a mutual interaction.
    Black dotted vertical lines indicate the point of intersection of the black lines, i.e., the resonance condition $\omega_\mathrm{wave} / \Omega_i = - \gamma_\crrm^{-1} $, at which we locally expect maximum growth. 
    However, the interaction with CRs modifies the induced wave frequency by $\delta \omega$, as indicated by red circles, leading to a modification of the resonance condition $(\omega_\mathrm{wave}+\delta\omega)/ \Omega_i = - \gamma_\crrm^{-1}$. Please refer to  Figure~\ref{fig:disp_bg} for a representation of this solution in the background frame.
	}
	\label{fig:disp_CRframe}
\end{figure*}

\subsection{Revisiting the dispersion relation}
As argued in the preceding subsection, our model predicts that the induced wave velocity is slower than the pristine wave velocity without streaming CRs. 
In the following, we investigate whether this finding is also manifested in the dispersion relation.
The dispersion relation is given in Appendix~\ref{sec:DR_formulae}.
We choose the parameters of the simulation \Isim to visualize the solution of the dispersion relation, which includes the backward and forward moving Alfv\'en wave (which are not resolved in the simulation \Isim) and the excited whistler wave from the intermediate-scale instability. 
This allows us to showcase all relevant instabilities, and the physical interpretations are transferable to the \Gsims~simulations.

In  Figure~\ref{fig:disp_CRframe}, we show the wave rotation rates as derived from the dispersion relation for a gyrotropic distribution of CRs, and evaluate them in the comoving CR frame where $\vdr=0$ (i.e., not in the wave frame). Note that we show the same solution in the background frame in Appendix~\ref{app:disp_bg_frame} for convenience. 
In the comoving frame, the resonance condition is given by 
\begin{equation}    
\mathcal{R}(\omega, k) =k\vdr-\omega -\Omega_\crrm=  - \omega -\Omega_\crrm = 0.
\label{eq:resonance_CR}
\end{equation}
Indeed, this condition locates the waves with the maximum growth rate using the pristine rotation rate of the wave, i.e., without taking into account mutual interaction (see the vertical dotted lines in Figure~\ref{fig:disp_CRframe}).

However, there are more subtleties related to the interplay of the CRs with the background modes that go beyond this simplistic view. The color of an individual solution to the dispersion relation shown in Figure~\ref{fig:disp_CRframe} describes a singly-connected branch that may or may not change character as it crosses a resonance. As we move from the left (small $k$ values) to the right, we first see a degenerate branch (shown at small $k$ values in red and green) which rotates at $-\Omega_\crrm$ in the frame comoving with CRs. These solutions can be interpreted as CR ion cyclotron wave modes \citep{Shalaby2023}. This solution splits up at $k d_i\approx 0.1$ as a result of the interaction of CRs with the backward moving Alfv\'en wave. At $k d_i\approx 0.14$ the upper CR ion cyclotron wave (green) interacts with the backward Alfv\'en wave (orange) so that the rotation rates exactly overlap, implying that their wave frequencies become degenerate. These degenerate waves complement each other, as their growth rates correspond to $\pm \Gamma$ of which only the positive part is shown in gold in the zoom-in panels. This describes a transfer of energy from one degenerate wave mode to the other, which implies an instability. As this degenerate solution of CR-backward Alfv\'en waves approaches the scales of forward Alfv\'en wave (initially denoted in blue), there is again energy exchanged between CRs and the background that changes the character of this particular wave and causes it to turn into a faster-rotating CR ion-cyclotron wave (CR branch 1). At the same time, the degenerate solution splits up into a new forward moving Alfv\'en wave (green) and a slower rotating CR ion-cyclotron wave (orange, CR branch 2), which approaches the upper CR wave at even smaller scales (larger $k$ values). We observe a similar behavior as we approach the resonance at the intermediate scale, where the interaction of CR ion cyclotron waves (blue and orange) with whistler waves (green) causes the rotation rates of the CR modes to deviate from each other so that the slower rotating CR mode (orange) overlaps with the modified whistler branch (green) and becomes degenerate, thus enabling the intermediate-scale instability. At smaller scales, the upper CR branch (blue) turns into a pure whistler wave.

Interestingly, a true degeneracy of a solution that either represents a CR ion cyclotron and a background branch or two CR branches only occurs provided the CR rotation rate is in between two (modified) background modes. This degeneracy gives rise to instability and is realized in between the forward and backward Alfv\'en modes for the CR streaming instability as well as in between the whistler and electron cyclotron modes for the intermediate-scale instability, see also Figure~\ref{fig:overviewInstability}. Note that the growth rates are still maximized close to the resonances and significantly reduced in between the background modes, where the unstable solution represents two CR branches.
This can also be seen in the lower left panel of Figure~\ref{fig:disp_CRframe}, where the pristine backward and forward moving Alfv\'en waves are shown in black, moving at $-\valf$ and $+\valf$. 
Interactions with the CR branch modifies their rotation rates and causes unstable waves with speeds in between this range, $-\valf\lesssim\varv\lesssim+\valf$.
This enables wave growth at these smaller velocities, even at low CR densities.
At scales larger than the ion skin depth, we often refer to these low-velocity CR-driven waves as Alfv\'en waves, which carries the connotation that they should propagate at $\pm\valf$, even though it would be more precise to characterize them as combined Alfv\'en-CR branch waves instead, as explained above.

We now focus on the fastest growing wave modes at resonance and its frequency shift $\delta\omega$.
The resonances are marked by the vertical gray dotted lines, which intersect the dispersion relation at $\delta\omega$, as marked by a red circle.
Indeed, we find that the frequency shift counteracts the wave speed, that is $\delta \omega>0$ for the backward moving wave (left-most resonance) while $\delta\omega<0$ for the two forward moving waves.
This is in line with our model, which predicts that $\delta\omega$ opposes the wave speed, and explains why the CRs bunch up on average at $\varphi<0$ for the backward moving wave and at $\varphi>0$ for the forward moving wave. Note that this has been found to be a necessary condition for instability.
In conclusion, the unperturbed resonance condition can be used to estimate the wave number $k$ with maximum growth. 
The actual observed resonance is perturbed by a small rotation rate, which is required for a positive feedback loop and, hence, for instability.

That is, for all unstable resonantly driven wave modes, the resonance is predicted by equating the isolated wave rotation and the Doppler-shifted CR ion-cyclotron wave mode $\omega = k \vdr - \Omega_\crrm $ \citep{Shalaby2023}.
However, the inclusion of CRs modifies 
the wave rotation rate $\omega$ by $\delta \omega$ at resonance. 
Therefore, the resonance condition is altered into $\omega+\delta \omega = k \vdr - \Omega_\crrm = k \varv_\parallel (0) - \Omega_\crrm$, where the last equality only holds for the CR distribution we consider in this work and $\delta \omega <0 ~ (> 0)$ for forward (backward) propagating wave modes at resonance.
$\delta \omega$ is counteracting the propagation direction and therefore slowing the wave down, which changes the wave frame, as $\vwa=(\omega+\delta\omega)/k$.
That is, the CR velocity in the wave frame, which has been used in equation~\eqref{eq:CRdphidt}, can be transformed into the background frame via $ k \vparwave(t) =  k \varv_\parallel - \omega - \delta \omega$, and thus, equation~\eqref{eq:CRdphidt} can be written as
\begin{equation}
\dot{\varphi}(t) =  k \Delta  \varv_\parallel - \delta \omega,
	\label{eq:phi_full}
 \end{equation}
where we define $ \Delta \varv_\parallel(t) \equiv \varv_\parallel(t) -  \vdr(0)$, and used the resonance condition to set $k\vdr-\omega-\Omega_\crrm=0$.

\begin{figure*}
	\centering
		\centering  \Large\Gfsim \par
		\plotone{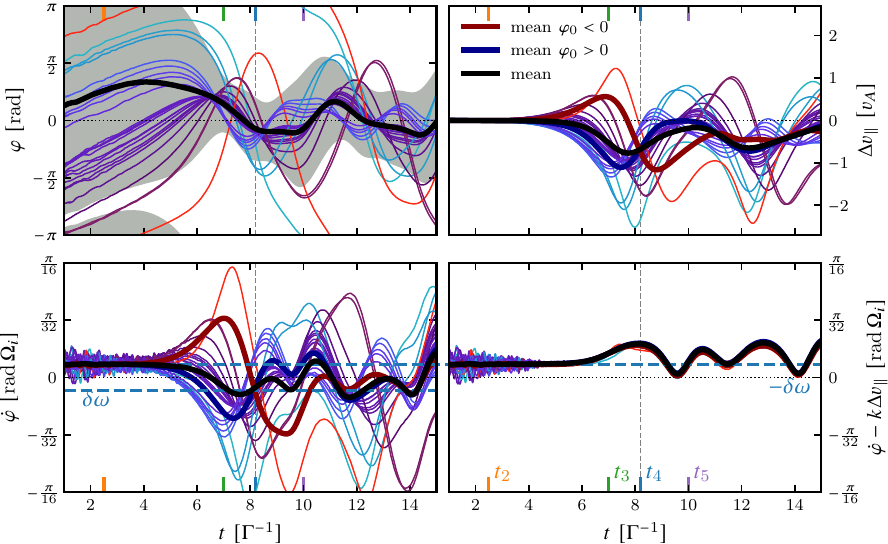}
		\par
		\centering \Large\Isim\par
		\plotone{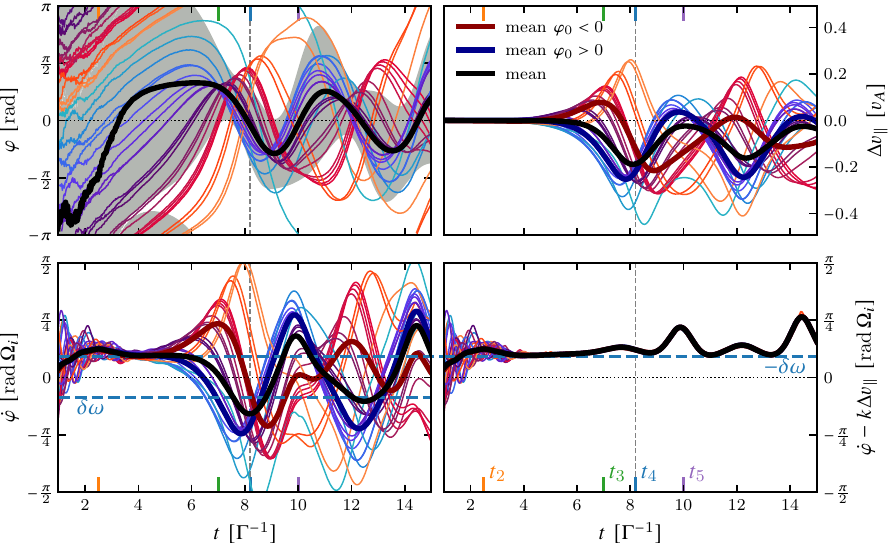}
	\caption{Simulated CR orbit parameters, which support our theoretical considerations regarding instability growth and the associated asymmetric bunching.
    We compare the cases of exciting forward propagating Alfv\'en waves (\Gfsim, top 4 panels) and whistler waves via the intermediate-scale instability (\Isim, bottom 4 panels). 
    For each simulation, we show the evolution of $\varphi(t)$, $\varv_\parallel(t)$ in the wave frame, the time derivative $\dot{\varphi}(t)$ and $\dot{\varphi}-k\Delta\varv_\parallel$, which is a measure of the wave rotation rate relative to the particle frame.
	Each panel shows multiple CR particles from a simulation with a single gyroresonant wave mode, mean values (thick black lines) and circular variance (gray band) are computed from 500 particles at random positions in the simulation box.
	The particle trajectories are colored from red to blue based on a gyrophase $\varphi_0 = \varphi(t=6)$, i.e., one $e$-fold before entering the nonlinear phase. Mean values for the different populations $\varphi_0>0$ and $\varphi_0<0$ are shown in corresponding colors. This analysis clearly demonstrates asymmetric CR bunching in $\varphi>0$ and confirms our theoretical picture that the CRs experience the unstable wave at relative rotation speed of $\delta\omega \approx \omega_\varphi-k \varv_\parallel$ at resonance.}
	\label{fig:phi_sim}
\end{figure*}

\subsection{Simulated family of particle orbits}
Figure~\ref{fig:phi_sim} enables us to test our predictions for the linear growth regime. This figure compares various CR orbit parameters for the CR streaming instability, which excites forward moving Alfv\'en waves (top 4 panels), and the intermediate-scale instability at the whistler scale (bottom 4 panels). For each instability, we show (1) the gyrophases, $\varphi(t)$, of a representative sample of CRs in the wave frame, as well as the mean of a large random sample of CRs, $\varphi_\mathrm{av}(t)$, (2) the parallel CR velocity in the wave frame, $\varv_\parallel(t)$, as well as the mean of the large CR sample, $\varv_{\parallel,\mathrm{av}}$, (3) the time derivative of the CR gyrophase $\dot{\varphi}(t)$, and (4) the quantity $\dot{\varphi}-k\Delta\varv_\parallel$, which is a measure of the wave rotation rate relative to the CR frame. During linear growth (for $t<t_3$), this analysis confirms our analytic predictions and supports the physics underlying the feedback loop described above.
First, there is an anisotropy developing in the CR gyrophases with $\varphi_\mathrm{av}>0$. Second, particles are accelerated along $B_0$ such that there is a net momentum loss of CR ($\Delta\varv_{\parallel,\mathrm{av}}<0$). Third, the particles' angular velocity evolution is asymmetric for the subpopulations with $\varphi_0 >0$ and $\varphi_0<0$, respectively (where we select $\varphi_0 = \varphi(t=6\Gamma^{-1})$ towards the end of the linear growth phase for visualization purposes). Fourth, there is a universal, almost constant frequency shift $ \delta \omega$ observed due to the violation of the resonance condition.

We now detail the evolution of the CR gyrophases. The fact that the CRs' gyrophases are not randomly distributed but instead show a coherent bunching over time is due to the resonance between CRs and the wave.
CRs with the same gyrophase $\varphi_0$ develop similarly, and $\varphi(t)$ of all particles shows a similar slope during the linear growth phase.   
To visualize the bunching of the CR gyrophases over time, we show the circular variance of $\varphi(t)$ (multiplied by $2\pi$) with a gray band that is centered on $\varphi_\mathrm{av}(t)$ (shown in black).  
The circular variance is also directly connected to the perpendicular CR current, $J_{\crrm,\perp}$, and the bunching in $\varphi$ causes an increasing CR current.
While the initial value of $\varphi_\mathrm{av}$ is noisy and physically irrelevant, it becomes decidedly positive for the forward moving waves as the particles bunch up towards $\varphi_\mathrm{av}>0$.
After saturation, most particles oscillate around $\varphi=0$, which indicates that these are trapped in the potential well.
However, some CR trajectories observed in the plot swing over and take on more complicated trajectories.

In each of the two cases, the top right panels show that particles starting off at $\varphi_0<0$ are accelerated in the parallel direction. Because this is aligned with the direction of the propagating wave, these CRs take momentum from the wave.
The decelerating particles ($\varphi_0>0$) are more numerous and thus, there is a net momentum gain by the wave at the expense of the CRs.

The bottom left panels show the time derivatives of the CR gyrophases, i.e., the instantaneous slopes of the CR trajectories shown in the top left panels. 
Once the wave mode starts to dominate the noise, $\dot{\varphi}$ remains almost uniform in the linear phase, which corresponds to the similar slopes of $\varphi(t)$ in the upper left panel. The similarity of the CR trajectories in the bottom left and top right panels shows that the particle acceleration term $k\Delta\varv_\parallel$ dominates the evolution of $\dot\varphi$ in equation~\eqref{eq:phi_full}. 

We use equation~\eqref{eq:phi_full} to estimate the resonance condition in the bottom right panel, 
\begin{align}
	\mathcal{R}(\omega, k)&=\dot{\varphi}-k  \Delta \varv_\parallel = -\delta \omega \quad \text{at resonance}.
\end{align}
$\delta \omega $ is the frequency shift introduced by the CRs, and the theoretically expected value obtained from the above dispersion relation is added as a dashed blue line in Figure~\ref{fig:phi_sim}. 
The theoretical expectations for $\delta\omega$ are clearly very similar to the rotation rate in the resonance frame in the linear growth phase.
In the nonlinear phase, the mean of $\delta\omega$ also oscillates due to changes in wave velocity, that is, $\delta \omega(t)$ is not constant in this phase.
This is discussed further in Section~\ref{sec:nl_wavevelocity}.
Deviations from the mean by individual particles are small, indicating that this is indeed a modification of the wave affecting all CRs at the same time. 
However, some individual deviations from the mean can be observed in \Gfsim. 
These are due to the direct Lorentz-force term along $\vec\varv_\parallel' \cross\vec{B}_\perp$ acting on $\psi_\crrm$ which has been neglected initially, see equation~\eqref{eq:lorentz_psi}.
Due to the larger wave velocity and slower $\vdr$, $\vparwave$ is significantly smaller in \Isim compared to $\vparwave$ in \Gfsim, which is why these deviations from the mean are more visible in the latter simulation.
Still, they are insignificant in comparison to the term $k \Delta \vparwave$ and thus, the neglect of this effect is justified. 

The nonlinear phase after $t_3$ is discussed in the next section.
Although the CR streaming and intermediate-scale instabilities in \Gfsim and \Isim act on very different spatial and temporal scales, the fundamental physical processes regulating wave growth are the same.
While we omit a similar plot of \Gbsim for brevity, the results are fundamentally similar, but qualitatively mirrored horizontally around $0$.

\begin{deluxetable*}{l|c|ccc|cc}
\tablecaption{Wave velocities and predicted and simulated pendulum frequencies.}
\label{tab:pendulum_frequency}
\tablehead{\colhead{Simulation} & \colhead{$\varv_\mathrm{wave} [\valf]$}&
\multicolumn{2}{c}{$\omega_\mathrm{pend} \left[\Omega_i\right]$}   & \colhead{Rel. error $\left[\%\right]$} & \multicolumn{2}{c}{$\omega_\mathrm{pend}\left[\Gamma\right]$}  \\
\colhead{} & \colhead{} & predicted & simulated & \colhead{} & predicted & simulated } 
\startdata
\Gbsim & -0.33 & 0.079 & 0.076 & 3.8 & 1.364 & 1.312 \\
\Gfsim & 0.72 & 0.100 & 0.083 & 17 & 1.562 &  1.297 \\
\Isim (\Iwsimm) & 4.73 & 0.723 & 0.736 & 1.8 & 1.482 & 1.508 \\
\enddata
\end{deluxetable*}

\section{Saturation of a single wave mode}
\label{sec:nonlinear}

During the linear growth phase, the CRs fall into the potential wells arising from the magnetic wave, see  Figure~\ref{fig:pendulum_potential}.
The nonlinear phase starts when an appreciable amount of CRs pass the minimum of the potential well and their acceleration direction reverses, that is roughly from $t_3$ onward.
In this section, we discuss three effects. First, the saturation time $t_4$ of the instability when the majority of CRs have passed the potential minimum, which is defined as the reversal point of the growth of $B_\perp$, after which it starts to decline; second, the modification of the unstable waves as a result of CR feedback, and finally, the anisotropy introduced into the CR distribution function.

\subsection{Pendulum oscillations during the nonlinear phase}
\label{sec:nonlinear_saturation}

Here, we investigate Figure~\ref{fig:phi_sim} with regard to the saturation time $t_4$, which is marked as a gray dashed line. 
At this time, the average $\varphi$ reverses, switching from $\varphi>0$ to $\varphi<0$, while $\dot{\varphi}$ has reached a local minimum. 
As described beforehand, this reverses the overall acceleration direction and leads to more particles taking momentum from the wave, thus damping it according to the drawing of Figure~\ref{fig:pendulum_schematic}.  
The perturbation of the wave frequency, which has been determined from the dispersion relationship beforehand,  has been identified as the rate, with which the potential wells move in relation to the particles.
Naturally, as the particles accelerate in parallel direction, they overtake the potential wells. 

In the linear phase, the potential wells move away from the particles at a relative velocity of $\delta \omega/k$, as can be seen from the bottom-right panel in Figure~\ref{fig:phi_sim}.
As a necessary condition, the particles need to move faster than the potential well to catch up.
At $t_4$ we observe that $\dot{\varphi} \sim \delta \omega$, and thus, using equation~\eqref{eq:phi_full}, the CRs move approximately twice as fast as the potential well with $\Delta \vdr \sim 2 \delta \omega/k$, where we adopted the mean over the particle distribution.
Inserting this estimate for $\Delta \vdr$ into the momentum equation~\eqref{eq:CRmomentumeq} yields a rough estimate for the saturation level of the magnetic wave field,
\begin{equation}
\frac{\Delta B_\perp^2}{B_0^2} \sim -\frac{n_\crrm}{n_\mathrm{bg}} \frac{\bar\gamma\vwa}{\valf^2}  \frac{2 \delta\omega}{k}. 
\end{equation}
This estimate is of the same order as the measured saturation level, e.g., for the \Isim~simulation we find that $B_\perp = 0.0131 B_0$ at saturation, while we predict $B_\perp = 0.0076 B_0$. 

\citet{sudanTheoryTriggeredVLF1971} proposed that the wave should saturate once the pendulum frequency, obtained from equation~\eqref{eq:CRpendulum} for small angles of $\varphi$, is comparable to the growth rate:
\begin{align}
    	\Gamma \sim \omega_{\rm pend}&=\sqrt{\gamma k_\mathrm{res} d_i \frac{\vperp}{\valf} \frac{B_\perp}{B_0}} \Omega_\crrm 
	\label{eq:GammaPendulum}\\
 \Leftrightarrow\quad
 \frac{B_\perp}{B_0} &\sim \left(\gamma k_\mathrm{res} d_i\right)^{-1} \frac{\valf}{\vperp} \left(\frac{\Gamma}{\Omega_\crrm}\right)^2. 
	\label{eq:BsatPendulum}
\end{align}
Hence, the unstable wavelength at resonance as seen from a gyrating relativistic CR appears to be Lorentz contracted and -- in tandem with the wave growth rate and $\varv_\perp$ -- determines the saturated magnetic wave field.

The pendulum frequency can be readily obtained from the oscillations of the magnetic field strength in  Figure~\ref{fig:bgrowth}. 
We thus start by estimating the simulated pendulum frequency using a least-squares fit of the magnetic wave amplitude, while calculating the theoretical pendulum frequency using the mean $\vperp$ and $B_\perp$ values from the simulation data after saturation ($t>t_4$).
The pendulum frequencies are given in Table~\ref{tab:pendulum_frequency}.
The excellent agreement of the simulated oscillations of the wave magnetic field and our theoretical estimates support our picture that CRs collectively behave as a pendulum in the wave magnetic field.
Upon closer inspection, the oscillation frequencies in the \Gsims~simulations are not perfectly sinusoidal (Figure~\ref{fig:bgrowth}), as they have longer growth phases and shorter damping phases, which could explain the larger relative error in these simulations. 

Next, we would like to scrutinize whether the growth rate $\Gamma$ is comparable to $\omega_{\rm pend}$ (equation~\ref{eq:GammaPendulum}).
These ratios are given in the two rightmost columns of Table~\ref{tab:pendulum_frequency}, and are similar enough to indicate that the oscillation frequencies of all gyroresonant instabilities at saturation are related to their growth rate, as predicted by equation~\eqref{eq:GammaPendulum}.
A parameter scan of different $n_\crrm/n_\mathrm{bg}$ for the forward moving Alfv\'en wave conducted by \citet{holcombGrowthSaturationGyroresonant2019} also supports our theory.

Although the growth rate of \Isim is larger than that of \Gsims, this does not necessarily imply that the intermediate-scale instability also dominates at saturation.
This is because it excites waves at a smaller scale (larger $k_\mathrm{res}$) and thus saturates earlier, according to equation~\eqref{eq:BsatPendulum}, leading to similar saturation levels between all of our simulations.
Instabilities at even smaller scales, like the electron cyclotron wave \Iesimm, can have larger physical growth rates in comparison to the other instabilities, but the saturation level is still expected to be substantially lower. This is in particular the case for a large scale separation between the unstable scales of \Iwsimm and \Iesimm, which is realized for small values of $\varv_\mathrm{dr}/\valf$ \citep{shalabyNewCosmicRaydriven2020}. 
If our simulations exactly resolved the peak of \Iesimm, we would expect a saturation rate of $B_\perp/B_0=2.0\times10^{-4}$, which is roughly $1/70$ of the saturation level of \Iwsimm. However, because of our discrete sampling, we expect a reduced growth rate for \Iesimm (as explained in Section~\ref{sec:setup}) so that this instability does not significantly influence the \Isim~simulation.

\begin{figure*}
   \resizebox{\hsize}{!} { \includegraphics[width=\textwidth]{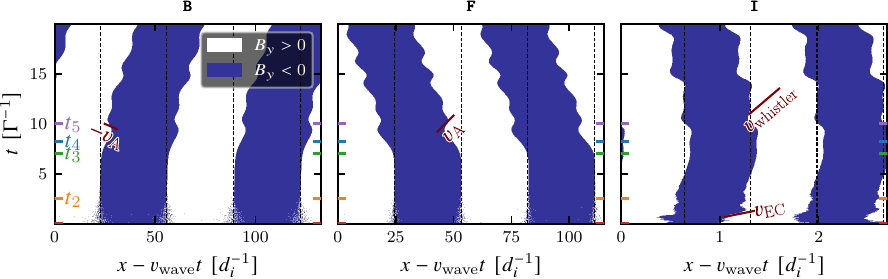} }
	\caption{
		Visualization of the time evolution of the wave velocities of the resonant modes in the simulations \Gbsim, \Gfsim, and \Isim. The $x$-axis of the plot is comoving with the wave at the speed $\vwa$, which is calculated from the dispersion relation at resonance while taking into account the CR perturbation: $\vwa/\valf=[-0.33, 0.72,4.73]$ from left to right (where we adopted the \Iwsimm wave speed in the latter case). 
		Dashed vertical lines indicate waves moving at this constant velocity.
		The slopes of the boundaries, where $B_y$ changes sign, are a useful visual indication of the wave velocity.
		The wave velocity nearly approaches the pristine (unmodified) wave velocity in the nonlinear phase in the first two panels, i.e., $\mp \valf$. 
        The corresponding whistler and electron cyclotron velocities $\varv_\mathrm{EC}$ are indicated in the third panel.
		The times $t_{1\text{--}5}$ are marked with their corresponding colors on the $y$-axis.
	}
	\label{fig:wavespeed}
\end{figure*}

\subsection{Impact of CRs on the wave velocity}
\label{sec:nl_wavevelocity}
In this paragraph, we investigate the CR back-reaction on the unstable waves. 
In the linear phase, the unstable waves behave as predicted by the dispersion relation while the wave speeds in the nonlinear phase show a more complicated behavior.
As the particles swing back and forth in the nonlinear wave magnetic field (see the bottom right panel of  Figure~\ref{fig:phi_sim}), this oscillating CR current directly impacts wave propagation. 
These perturbations in $\delta \omega$ over time indeed affect the wave velocity, which is shown in Figure~\ref{fig:wavespeed}. 
In all three cases after $t_4$, the absolute wave velocities slow down on average in comparison to their propagation speeds during linear growth.

The influence of CRs on the wave speed is substantially underestimated by adopting the formal definition of the Alfv\'en speed, $\valf=B_0/\sqrt{\mu_0 \sum_s m_s (n_s+n_\crrm)}\approx B_0/\sqrt{\mu_0 \sum_s m_s n_s}$ (for $n_\crrm/n_\mathrm{bg}\ll1$), which weights CRs only by their comparably small mass density. 
This is because the equation for the Alfv\'en velocity assumes a plasma at rest.
However, the current of CRs, $\vec{J}_{\crrm,\perp}$, is generated as a result of the magnetic bunching process, with an amplitude comparable to the background current, $\vec{J}_{\mathrm{bg},\perp}$.
In consequence, the unstable waves in the \Gsims~simulations propagate at $-0.33\valf$ and $0.72\valf$, significantly slower than the unmodified Alfv\'en speed. 

However, the unstable waves reach their corresponding (unmodified) Alfv\'en speeds at the rebound point $t_5$ in the fully nonlinear phase.
Hence, in order to estimate the saturation level of the instability, one would have to use the corresponding wave velocity in the momentum equation~\eqref{eq:CRmomentumeq}, i.e., the modified wave velocity. 
As the wave velocity is already varying before saturation, we would have to take into account those changes over time in the momentum equation, which is not trivial.

The case of the \Isim~simulation is even more complex. As the unstable \Iesimm wave saturates shortly after $t_2$, there are several modes excited with wave velocities in between the \Iwsimm and the \Iesimm modes (see Figure~\ref{fig:wavespeed}). After that time, there is a spectrum of waves close to the \Iwsimm resonance excited. The resulting combined wave field propagates at a velocity that is somewhat faster than expected for the purely growing \Iwsimm wave. In the saturated stage after $t_4$, this wave is considerably slowed down in response to the oscillating CR current.

\subsection{Evolving CR distribution}

In the following, we examine changes to the CR distribution as a result of their wave-particle interactions.
As the gyrophases of the CR ions generally follow the pendulum equation~\eqref{eq:CRpendulum}, their evolution mostly depends on the initial conditions $\varphi_0$ and $\dot\varphi_0$.
In  Figure~\ref{fig:phi_sim}, we defined $\varphi_0=\varphi(t=6)$ shortly before the nonlinear phase, which is when $\dot\varphi_0$ is still comparably small and thus negligible in the initial conditions. 
Because more particles have $\varphi_0>0$ (which is the necessary condition for the instability to grow), these dominate the overall mean. 
A difference in the mean values of the $\varphi_0>0$ and $\varphi_0<0$ populations is indeed observed.
Most notably, these swing out of phase with each other in the nonlinear regime, as expected from evolving pendulum with out-of-phase initial conditions.

\begin{figure}
   \resizebox{\hsize}{!} { \includegraphics[width=\textwidth]{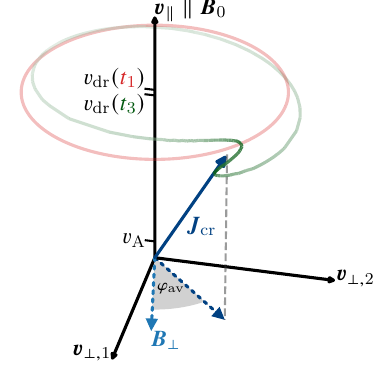} }
	\caption{A visualization of the velocity space of simulation \Gfsim. The CR ions have changed from their initial distribution at $t_1$ (shown in red), to a bunched up, non-gyrotropic distribution at the onset of the nonlinear phase (at $t_3$, shown in green; denser areas are depicted by darker colors).
	All CR angles are measured in $\varphi$ instead of $\psi$, which captures the natural anisotropy in the perpendicular plane introduced by the $\vec{B}_\perp$. }
	\label{fig:gyro_conclusion} 
\end{figure}

Figure~\ref{fig:gyro_conclusion} shows the CR velocity distribution of \Gfsim at the onset of the nonlinear phase $t_3$.
At this time, most particles share a similar $\varphi$ (as a result of the bunching), while the rotational velocity $\dot{\varphi}$ is notably different. 
Particles with $\varphi_0<0$ have been accelerated in parallel direction (i.e., their velocities depicted in green lie above their initial values, shown in red), while those with $\varphi_0>0$ have a smaller $\varv_\parallel$ in comparison to the initial gyrotropic ring distribution.
Because particles are still accelerated by the magnetic field, this state is only quasi-stable -- from $t_3$ onwards, the particles oscillate around the wave magnetic field, $\vec{B}_\perp$. 
Therefore, they cyclically bunch up and spread out again, even though most particles stay roughly aligned with $\vec{B}_\perp$.

\section{Conclusions}
\label{sec:conclusions}

In this work, we studied the physics of streaming CRs in a background magnetic field and the associated excitation of plasma instabilities from first principles. We developed a theory of the underlying processes that organize the particles' orbits and in particular their gyrophases, which provides an intuitive physical picture of the growth, saturation, and back-reaction onto the plasma waves excited via CR-driven instabilities. However, for transparency, we restrict ourselves to single unstable modes. Starting from a gyrotropic setup of CRs, which embraces the symmetry of a magnetized plasma, we find that resonantly driven electromagnetic waves introduce an additional asymmetry perpendicular to the background magnetic field. As a result, a new stable equilibrium state emerges as the gyrophase of the CR ions follows this asymmetry to locally match the phase of the driven waves.

Based on our simulation results and theoretical considerations, our new theory for the growth of individual unstable waves driven by CRs contains the following elements. 
\begin{itemize}
	\item Wave growth resulting from an instability necessitates $k>0$, but occurs independent of the wave propagation direction. 
    Fast moving CRs with $\varv_\parallel>\valf$ can excite forward and backward-propagating Alfv\'en waves (as measured in the background frame). 
	\item The unstable waves cause the CRs to bunch up in gyrophase through parallel Lorentz forces. In result, the CR distribution develops a lopsided gyrophase with respect to the local wave magnetic field. In other words, the helical wave magnetic field is joined by a helical CR distribution (i.e., a CR ion cyclotron wave) that exhibits the exact same winding angle as the unstable wave. This lopsidedness is key for enabling momentum transfer from the CRs to the wave and thus, for instability.
 	\item CRs are scattered asymmetrically parallel to the background magnetic field, preferentially but not exclusively in the direction opposite to that of the propagating waves. This is a secular scattering process dictated by the direction of the parallel Lorentz force and not a diffusive scattering process.
 	\item CRs modify the wave velocity, which is always slower (in absolute terms) than the wave velocity without CRs. This effect is especially pronounced for induced Alfv\'en waves, which propagate at velocities significantly less than $\valf$.
 	\item The instability saturates once the majority of CRs become fast enough to overtake the unstable wave, which propagates at the CR-modified wave speed. In consequence, the wave is slowed down by the faster CRs, which implies wave damping and hence saturation of the instability.
    \item The motion of the (bunched up) CRs relative to the local wave magnetic field can be described by a pendulum equation. In this picture, linear wave growth of the instability results from the CR approaching the local wave magnetic field. As CRs overshoot the equilibrium position of an exact alignment of CRs and the local wave field, the instability saturates. The nonlinear behavior of the instability is then characterized by an oscillating CR distribution in the potential associated with the parallel Lorentz force, which is centered on the local wave field. This oscillating CR distribution generates perpendicular CR currents, which also cause the wave amplitude to oscillate and to further slow down.
\end{itemize}

Additionally, we find that the three main instabilities, which describe resonant interactions of streaming CRs with forward and backward traveling Alfv\'en waves, as well as with whistler waves via the intermediate-scale instability (if present) saturate via the exact same mechanism and to similar amplitudes in our setup. 
Unstable waves on smaller scales, such as the CR-driven electron cyclotron wave of the intermediate-scale instability, saturate to lower levels, even though their growth rate might dominate.

In this work, we did not fully explore the relative importance of these three instabilities and instead concentrated on studying their underlying physics while adopting the simplest possible configuration for transparency. Thus, this paper is meant to  provide a starting point for future research. Of prime importance will be the study of the differences of the CR-wave scattering for these various instabilities as this directly impacts the effective CR transport speed and momentum transferred by the CRs to the background plasma. Second, we need to extend the theory developed here for the growth of isolated wave modes to include wave-wave interactions of the unstable modes of the forward and backwards Alfv\'en and whistler branches, which could yield a modification of the criterion for instability saturation. Third, a necessary extension of this work would also be to generalize the initial gyrotropic ring distribution of CRs to a more natural power-law momentum distribution exhibiting all CR pitch angles.

While wave growth induced by a power-law momentum distribution of CRs is expected to be different from that of single wave modes, we believe that some of our main results such as CRs bunching up in gyrophase as a requirement for driving the instability by means of this anisotropy will carry over. Regardless, the results shown here indicate that some of the general assumptions commonly applied to CR transport based on quasi-linear theory could be violated.
This includes the random phase approximation, as the gyrophase of CRs is potentially strongly correlated with the driven waves.
Furthermore, our results indicate that the saturation level may not be estimated from the momentum equation~\eqref{eq:CRmomentumeq} by assuming that $\vdr$ asymptotically converges to $\valf$ as it isotropizes in the frame of the forward moving Alfv\'en wave.
Instead, the growth of different, potentially important wave modes with temporally changing wave velocities makes estimates using the momentum equation difficult and could identify the erroneous isotropization frame.
Running physically motivated simulations requires great care, as the results can radically differ according to the box size or mass ratio, as shown by the simulations presented here, which are opening the door for a rich avenue of future research.

\section{acknowledgments}
We acknowledge the support of the European Research Council under the ERC-AdG grant PICOGAL-101019746. This work was supported by the North-German Supercomputing Alliance (HLRN) under projects bbp00046 and bbp00072. This work was finalized at the Aspen Center for Physics, which is supported by the National Science Foundation grant PHY-2210452.
\newpage

\appendix
\section{Conventions}
\label{app:convention}
It is convenient to use a definition of vectors, which naturally reproduces the symmetry of the problem. 
To this end, we align the $x$ coordinate of our coordinate system with the background magnetic field $\vec{B}_0$ and denote parallel vector components as $\vec{b}_\parallel= b_x \vec{e}_x$, where $\vec{e}_x$ denotes the unit vector. We express the perpendicular plane in complex notation: adopting $\vec{e}_\perp\equiv \vec{e}_y$ and $\ci\,\vec{e}_\perp \equiv \vec{e}_z$  results in $\vec a_\perp=(a_y + \ci\,a_z)\vec{e}_\perp$. The following identity follows:
\begin{align}
	\vec{b}_\parallel \cross \vec{a}_\perp = -b_x a_z \vec{e}_y + b_x a_y\vec{e}_z = b_\parallel \ci\,\vec{a}_\perp.
\end{align}
Ions will experience a Lorentz force 
\begin{align}
	\vec{F} = q\vec{\varv}_\perp \cross \vec{B}_\parallel = -\ci\, qB_0 \vec{\varv}_\perp
\end{align}
around the mean field, which causes a gyration with $\vec{\varv}_\perp=\varv_\perp\vec{e}_\perp \exp(-\ci\,\Omega_i t)$. 

Transversal waves evolve in the perpendicular plane like $\exp\{\ci\,[(\omega -\ci\Gamma) t -k_\parallel x ]\}\vec{e_\perp}$, where $\omega$ and $\Gamma$ denote the wave rotation frequency and growth rate, respectively. According to this definition, waves with a positive $k_\parallel$ have a left-handed helicity for increasing $x$. Furthermore, the sign of $\omega$ indicates the polarization of the wave: $\omega>0$ corresponds to a right-handed wave while $\omega<0$ corresponds to a left-handed wave. The phase velocity of the wave, $\vwa=\omega/k$, is an invariant physical quantity and its sign indicates the direction of movement.

For reference, we compare our definition to another popular definition \citep{stixWavesPlasmas1992, baiMagnetohydrodynamicParticleincellSimulations2019}.  In the following, we denote quantities defined in that convention with tilde symbols. In this definition, $\tilde{\omega}$ is set to be always positive and our definition is recovered by setting $\omega=\pm \tilde{\omega}$, depending on the polarization of a wave. In their convection, the wave polarization cannot be inferred solely from $\tilde{\omega}$ but needs to be explicitly specified, i.e., right-handed (left-handed) polarized waves are corotating with the electrons (ions) and are denoted by $\exp[\ci\,(\pm\tilde\omega t -k_\parallel x )]\vec{e_\perp}$. Likewise, the definition of the wave velocity depends on the wave polarization: 
$\tilde\varv_\mathrm{wave}=\pm \tilde{\omega}/\tilde{k}$, where the plus (minus) sign corresponds to
a right-handed (left-handed) wave. Despite the difference in notation, both definitions for the wave velocity coincide, $\vwa = \tilde\varv_\mathrm{wave}$, because the wave velocity is a physical invariant.

\section{Comparison of fluid-PIC and PIC methods}
\label{app:PIC}

\begin{figure}
   \resizebox{\hsize}{!} { \includegraphics[width=\textwidth]{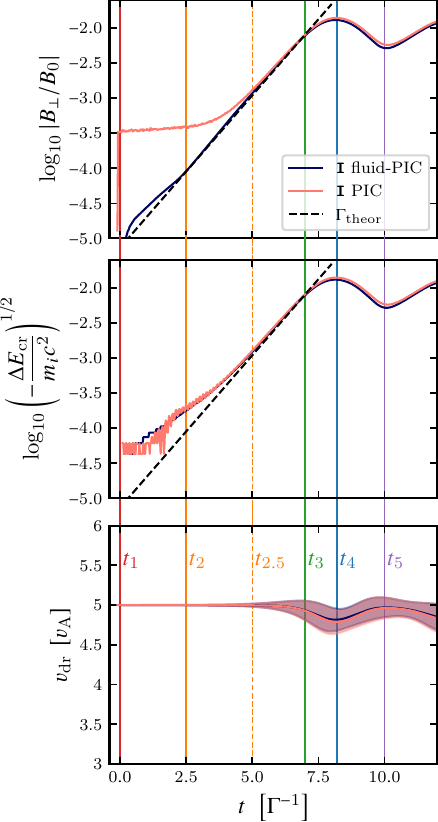} }
    \caption{
    Same as  Figure~\ref{fig:bgrowth}, but comparing \Isim~simulations performed using the fluid-PIC and pure PIC method.
    Both results are almost identical, except for the increased noise floor in the PIC simulation at early times.
    Due to the differences in the noise level, the time marker $t_{2}$, which initially denoted the linear growth phase, has been moved to a later time, $t_{2.5}$, to enable a fair comparison between the PIC and fluid-PIC methods in the linear growth regime.}    \label{fig:PIC-bgrowth}
\end{figure}
\begin{figure*}
    \plotone{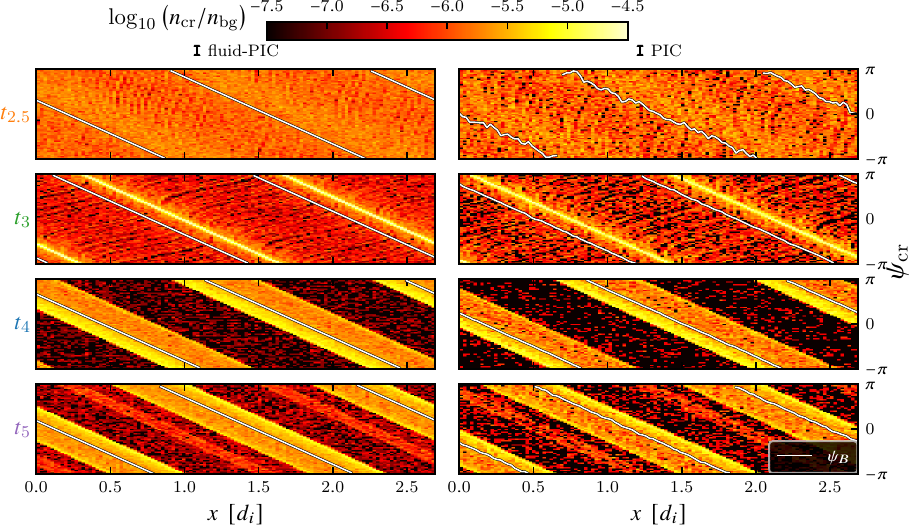}\caption{Same as  Figure~\ref{fig:xtscatter}, but comparing \Isim~simulations performed using the fluid-PIC (left panels) and pure PIC (right panels) methods.
    $t_{2.5}$ is used instead of $t_2$ to ensure a fair comparison between the two methods because of the increased electromagnetic noise floor in the PIC simulation at earlier times, see Figure~\ref{fig:PIC-bgrowth}. The CRs at time $t_{2.5}$ show more structure compared to $t_2$ used in Figure~\ref{fig:xtscatter} as a result of the larger magnetic wave intensity.}\label{fig:PIC-anglext}
\end{figure*}

The simulations in this paper have been performed using the novel fluid-PIC method \citep{lemmerzCouplingMultifluidDynamics2024}.
Here, we compare the \Isim~simulation obtained from this novel method with the traditional and well-tested PIC method using the SHARP code \citep{shalabySHARPSpatiallyHigherorder2017,shalabyNewCosmicRaydriven2020}.
We use the same parameters as defined in Section~\ref{sec:setup}, with the only exception that we use only $25$ computational particles per cell for the CR species and $2.5\times10^5$ particles per cell for the background species in the PIC simulations, in contrast to $75$ particles per cell used in the fluid-PIC simulations.
Because the fluid-PIC method does not have to follow particles of the background species, it is more computationally efficient.

The results for wave growth are shown in  Figure~\ref{fig:PIC-bgrowth}, which compares the growth of the unstable wave magnetic field as a result of the intermediate-scale instability (top panel), the energy loss experienced by the CRs in exciting this modified whistler wave (middle panel), and the mean drift speed of the CR population (bottom panel). Overall, both simulations produce nearly identical results except for the perpendicular magnetic field at early times, which shows an obvious difference in the noise floor of the simulations. 
This could be lowered in the PIC simulation by increasing the computational particles per cell, at the expense of becoming more computationally expensive.

In Figure~\ref{fig:PIC-anglext}, we compare the CR phases $\psi_\crrm$ and magnetic field angles $\psi_B$ for both simulations.
We can identify all important characteristics in both simulations: wave growth and the emergence of a helical magnetic field structure at $t_{2.5}$, asymmetric bunching of particles at $t_3$, saturation through particles oscillating at around $\vec{B}_\perp$ at $t_4$, and the back-swing as well as the \enquote{ghost} strip at $t_5$.
Because waves in both boxes grow from different realizations, the specific $\psi$ values are not expected to match in between different simulations, i.e., an offset in the phases of the waves is expected. 
As discussed before, the PIC results are noisier due to the smaller number of CR particles per cell. 
This also influences the wave magnetic field in the PIC simulation, which does not appear as straight white lines but instead shows wiggles as a consequence of small-scale noise generated by the shot noise of the finite CR and background particle number. 
Notwithstanding this minor difference, the physical effects described in this paper agree to high precision between both simulations and are thus independent of the choice of the numerical methods used.

\begin{figure*}
   \resizebox{\hsize}{!} { \includegraphics[width=\textwidth]{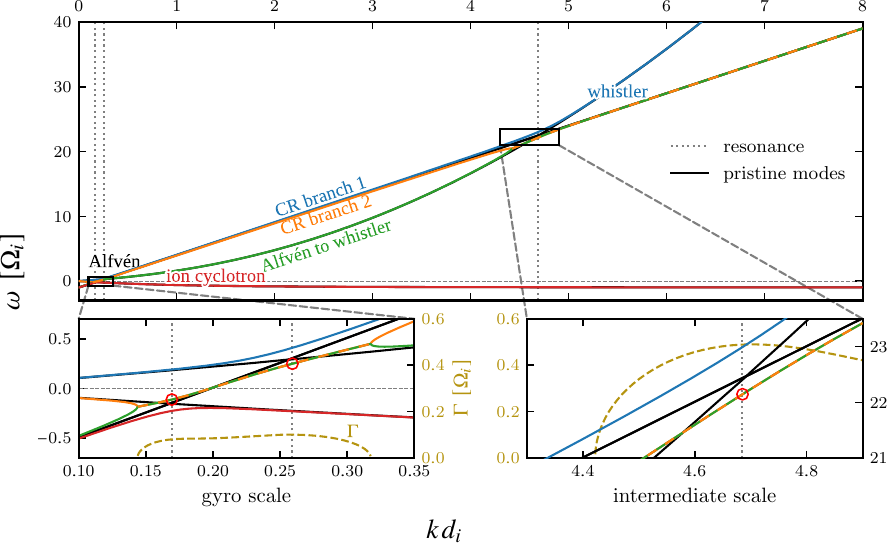} }
    \caption{The solution of the dispersion relation~\eqref{eq:fulldisp-bg} in the background frame using the parameters of the \Isim~simulation. Solid colored lines show the real frequency, while dashed colored lines show the growth rate of the wave modes. 
    Solid black lines indicate the dispersion relation of the pristine background modes and CR ion-cyclotron waves without taking into account a mutual interaction.
    Black dotted vertical lines indicate the point of intersection of the black lines, i.e., the resonance condition $\omega_\mathrm{wave} = k \vdr - \Omega_\crrm $, at which we locally expect maximum growth. 
    However, the interaction with CRs modifies the induced wave frequency by $\delta \omega$, as indicated by red circles, leading to a modification of the resonance condition $\omega_\mathrm{wave}+\delta\omega = k \vdr - \Omega_\crrm $.
    Please refer to Figure~\ref{fig:disp_CRframe} for a representation of this solution in the comoving CR frame. }
    \label{fig:disp_bg}
\end{figure*}

\section{Dispersion Relation}
\label{sec:DR}
\subsection{Equations}
\label{sec:DR_formulae}
We use the plasma dispersion function for transverse waves with a gyrotropic ring distribution \citep[e.g.][]{wuElectromagneticInstabilitiesProduced1972, Shalaby2023}
\begin{align}
	&\zeta_s ( \varv_{{\rm dr},s},\varv_{\perp,s}
	,n_s) = \nonumber\\
	&\quad\frac{ \omega^2_s }{ \gamma_s} 
	\left[
	\frac{\omega -k \varv_{{\rm dr},s} }{k \varv_{{\rm dr},s} -\omega - \Omega_s }
	-
	\frac{\varv_{\perp,s} c^{-2}\left(k^2c^2-\omega ^2  \right)}
	{2 \left(k \varv_{{\rm dr},s}  - \omega  - \Omega_s\right){}^2}
	\right].
	\label{eq:zeta}
\end{align}
Here, $\omega_{s} = \omega_{s}(n_{ s}) = \sqrt{ n_{ s} q^2_{s}/(m_{ s} \epsilon_0) }$ is the plasma frequency of a species $s$ and $\Omega_s$ is the relativistic cyclotron frequency for a particle of that species in the corresponding frame.
To obtain $\omega(k)$ in the frame comoving with the CRs, the following equation is solved numerically, and we account for the background species, CR ions and the neutralizing CR electron beam:
\begin{align}
	&\omega_{\mathrm{comoving}}^2- k^2 c^2
	+ \zeta_{\rm e} (-\vdr,0,  n_\mathrm{bg}) + \zeta_{\rm i} (-\vdr,0,n_\mathrm{bg})\nonumber\\
	&\quad
    + \zeta_{\rm e,  \crrm} (0,0, n_\crrm) + \zeta_{\rm i, \crrm} (0,\varv_{\perp, \crrm}, n_\crrm) = 0.
    \label{eq:fulldisp-CR}
\end{align}
Wave speed estimates in the background frame are retrieved from the following equation:
\begin{align}
	&\omega_{\mathrm{bg}}^2- k^2 c^2
	+ \zeta_{\rm e} (0,0,  n_\mathrm{bg}) + \zeta_{\rm i} (0,0,n_\mathrm{bg})\nonumber\\
	&\quad
    + \zeta_{\rm e, \rm \crrm} (\vdr,0, n_\crrm) + \zeta_{\rm i, \crrm} (\vdr,\varv_{\perp, \crrm}, n_\crrm) = 0.
	\label{eq:fulldisp-bg}
\end{align}
The pristine modes are recovered by solving these equations in the limit of $n_\crrm/n_\mathrm{bg}\rightarrow0$.

\subsection{Dispersion relation in the background frame}
\label{app:disp_bg_frame}

In Figure~\ref{fig:disp_bg}, we show the solution of the dispersion relation~\eqref{eq:fulldisp-bg} in the background frame using the parameters of the \Isim~simulation. This solution in the background frame shows the familiar behavior of the dispersion relation of the background modes, i.e., the backward and forward moving Alfv\'en wave ($\omega\approx \pm k\valf$) and the parallel whistler wave ($\omega\approx k^2 d_e^2\Omega_e$, where the electron skin depth is $d_e=c/\omega_e$ and the electron cyclotron frequency is $\Omega_e = qB_0/m_e$). In this frame, the changes of the character of a wave from pure background modes to degenerate CR-background modes due to CR-wave interactions are less obvious in comparison to the presentation in the comoving CR frame (Figure~\ref{fig:disp_CRframe}).

\bibliography{lib}
\end{document}